\documentclass[aps, pra, reprint, longbibliography,
superscriptaddress]{revtex4-1}

\usepackage{hyperref}
\usepackage{graphicx}
\usepackage[utf8]{inputenc}
\usepackage{xcolor}
\usepackage{amsmath, amssymb}
\usepackage{physics}
\usepackage{ulem}

\begin{document}

\preprint{APS/123-QED}

\title{Quasi-two-dimensional trapped tilted dipoles at zero and finite temperatures in the strongly dipolar regime}

\author{J. S\'anchez-Baena}
\email{juan.sanchez.baena@upc.edu}
\affiliation{Departament de F\'isica, Universitat Polit\`ecnica de Catalunya, Campus Nord B4-B5, 08034 Barcelona, Spain}

\date{\today}

\begin{abstract}

Motivated by the recent experimental observation of dipolar supersolid stripes in a quasi two-dimensional geometry [arXiv:2512.13280 (2025)], we study a trapped system of fully polarized dipoles in a strongly axially confined geometry, both at zero and finite temperatures, by means of Bogoliubov theory. The dipoles are strongly harmonically trapped along the $z$ axis and subjected to a box trap in the $x$-$y$ plane. We characterize the physics of the trapped system at zero and finite temperatures as a function of the tilting angle of the dipoles, the number of particles and the scattering length, restricting ourselves to the experimentally relevant regime of large condensate fractions. We also illustrate the influence of the aspect ratio of the box trap in the liquid character of the system and its structure. We observe a remarkable promotion of spatial modulations when temperature is increased while keeping the \textit{total} particle number constant for specific configurations, in qualitative agreement with previous Monte Carlo results in a 3D geometry.
Our results are useful to understand the zero temperature physics of the trapped dipolar system in the quasi-2D limit and in the strongly dipolar regime. In addition, they allow to assess the effect of temperature in its equilibrium properties in experimentally relevant conditions, which may be useful for thermometry applications.
\end{abstract}

\maketitle

\section{\label{sec:introduction}Introduction}

The interplay between the anisotropy of the dipole-dipole interaction (DDI) of an ultracold quantum gas and that of a confining potential gives rise to a rich scenario, with the main highlight being the formation of supersolid arrays of dipolar clusters. Supersolidity is a counter-intuitive phenomenon predicted more than fifty years ago~\cite{gross:1957} that combines spatial order with the frictionless flow of a superfluid. Its emergence in dipolar systems has sparkled an intense experimental activity in the recent years, primarily focused on three-dimensional systems~\cite{Modugno:PRL:2019,Pfau:PRX:2019,Ferlaino:PRX:2019,
Tanzi:Nature:2019,Guo:Nature:2019,Tanzi:Science:2021,norcia21:nature,
BiagioniPRX2022,Ferlaino:PRL:2021}. Recently, however, many efforts have been directed towards the realization of ultracold atomic systems in quasi-two dimensional (quasi-2D) configurations~\cite{sunami2022:prl,Yu2024:science,Guo2024:science,liao2026:arxiv}, including dipolar systems~\cite{he2025:scienceadv,zhen2025:arxiv,he2025:arxiv,he2026:arxiv}. In this setting, an axial harmonic trap is set such that the wave function of the atoms along the trapping axis can be approximated by the harmonic oscillator ground state. For dipolar atoms, this would allow to experimentally assess the interplay between the DDI and the effects of dimensionality, like the superfluid Berezinskii-Kosterlitz-Thouless (BKT) transition at finite temperature or the absence of true long-range order in the thermodynamic limit in two dimensions. As a result of these efforts, the BKT transition has been observed for a dipolar system in the weakly dipolar regime~\cite{he2025:scienceadv} and the effect of magnetostriction has also been observed and characterized for superfluid and thermal dipolar quasi-2D gases~\cite{he2026:arxiv}. In regards to supersolidity, supersolid stripes of dipolar bosons have also been observed in the quasi-2D regime for the first time very recently~\cite{he2025:arxiv}.

It is precisely the experiment of Ref.~\cite{he2025:arxiv} which drives the motivation of the present work, which also attempts to extend our previous analysis of the quasi-2D dipolar system at zero temperature and in the thermodynamic limit~\cite{baena2025:pra}. In the experimental set-up, the dipolar atoms are subjected to a finite temperature, even if it lies significantly below the BEC transition temperature. Recent theoretical and experimental work~\cite{baena22,baena24,He2024:PRR,bombin25:prl} has shown the remarkable importance of thermal effects in dipolar systems, which can promote spatial modulations and supersolidity. As such, in order to guide potential future experimental activity looking to further characterize the behaviour of dipoles in the quasi-2D regime, it is relevant to account for thermal effects, specially in the strongly dipolar regime, where the scattering length $a$ is smaller than the dipole length $a_{\rm dd}$ and supersolid structures arise.

Over the previous years, many theoretical works have explored dipolar systems confined in flattened and strictly 2D geometries~\cite{fischer_06_PRA, boudjemaa_2013_PRA, mishra_2016_PRA, Baillie_2015,fedorov_2014_PRA,mishra_2016_PRA,aleksandrova_2024_PRA,marchetti2013:prb,block2014:prb,lee_2024_PRA,amrey2015,soumik:PRA:2019,Boudjemaa2019:NJP,macia12b,macia12,macia14,bombin17:PRL,Sanchez-Baena_2018:NJP,guijarro22:PRL,pradas2022,staudinger23:PRA,guijarro2024,blakie2023:pra,poli2023:pra,cook2026:arxiv,matveekno2020:PRA,pinchenkova2024:arxiv,baena2025:pra}. Focusing on previous works considering finite temperature for dipolar bosons, c-field methods are employed in Ref.~\cite{pawlowski2013:PRA} to study the dipolar gas in the quasi-2D regime, which allow to obtain the statistical properties of the system. Ref.~\cite{pengtao2021} studies a pure 2D system of tilted dipoles in the thermodynamic limit by employing the Random Phase Approximation formalism (RPA) while Ref.~\cite{bombin19:PRA} does so via the Path Integral Monte Carlo method. Importantly, all of these methods do not rely on the existence of a Bose-Einstein Condensate. Bogoliubov theory has also been employed in the limit of ultralow temperatures for fully trapped systems~\cite{ticknor2012:PRA,ticknor2012:PRA2}. A similar configuration to the set-up of interest in this work is studied in Refs.~\cite{zhang19,zhang21,maucher24,He2024:PRR,cinti2025:PRA,soumyadeep2025:pra} at zero and finite temperature, where dipoles are also trapped along the $x$-$y$ plane but the system lies in the 3D limit, since the trapping along the $z$-axis is not strong enough to reach the quasi-2D condition. This distinction is specially relevant, since the dimensional character of the system has a huge impact on its properties, for instance, on the existence of a Bose-Einstein Condensate (BEC), as we discuss in Sec.~\ref{sec:theory_cond}.

In this work, we employ the Gross-Pitaevskii equation and Bogoliubov theory to study a system of tilted dipoles confined in a quasi-2D geometry, consisting of a strong harmonic trap along the $z$-axis and a box trap in the $x$-$y$ plane. We consider both zero and finite temperatures. The paper is organized as follows: in Sec.~\ref{sec:theory}, we discuss the existence of a BEC in our set-up and apply Bogoliubov theory to compute the thermal excitations for our specific geometry. In Sec.~\ref{sec:results} we present the results of our work while in Sec.~\ref{sec:conclusions} we summarize our conclusions.

\section{\label{sec:theory}Theory}

\subsection{\label{sec:theory_cond}Existence of a BEC in the quasi-2D regime}

It is well known that the Mermin-Wagner theorem forbids the existence of true long-range order in two dimensions. This has strong consequences in Bose-Einstein condensation, since this phenomenon is tied to the existence of off-diagonal long-range order in the one-body density matrix (OBDM). For systems in 2D, it is found that the OBDM $\rho({\bf r}, {\bf r'},)$ decays to a non-zero constant at zero temperature as $\abs{{\bf r} - {\bf r'}} \rightarrow \infty$ while it does so algebraically at any finite temperature below the BKT transition temperature $T_{\rm BKT}$, above which correlations decay exponentially to zero. Thus, it is generally stated that a BEC only exists in 2D and in the thermodynamic limit at zero temperature while it is destroyed by any thermal fluctuations. This fact is reflected in the calculation of the density of thermal atoms for a 2D ideal gas at finite temperature, which diverges for any $T \neq 0$. This remains true for an infinite, quasi-2D system strongly trapped along one direction (for instance, the $z$-axis) if $\hbar \omega \sim k_B T$, with $\omega$ the trap frequency, while the convergence of the thermal density is only recovered in the 3D limit ($\hbar \omega \ll k_B T$)~\cite{keepfer2022:PRA}.

Nevertheless, and crucially, the imposition of a trapping confinement in the infinite plane restores the convergence of the thermal atom number, and thus, the existence of a condensate~\cite{simulaPRA:2008,Bisset2008:PRA,Bisset2009:PRA,choi2013:prl}. The qualitative idea behind this is that the trap sets a maximum length scale at which off-diagonal long-range order has not vanished. This implies that, for cold enough temperatures, we can employ a formalism that assumes that there is a single quantum state macroscopically occupied (a BEC) that can be described by a Gross-Pitaevskii-like equation~\cite{ticknor2012:PRA,ticknor2012:PRA2}, while the fraction of thermally excited atoms remains small compared to the condensate. In all the calculations performed in this work, we restrict ourselves to the limit of large condensate fractions ($f_c \geq 0.8$). This means we work in the low temperature limit, away from the the trapped BEC transition temperature, $T \ll T_c$, and the BKT transition temperature, $T \ll T_{\rm BKT}$. In contrast, in the previous works described in the Introduction, which employ other finite temperature methods, the temperature is considerably closer to $T_{\rm BKT}$. Moreover, we focus entirely on the study of a trapped system, since in the infinite 2D limit the condensate is fully depleted by thermal effects and thus Bogoliubov theory is inapplicable. In the following Sections we derive this finite temperature GPE equation (which we refer to as TeGPE) by adapting the theory presented in Refs.~\cite{oktel19,oktel20,baena22} to our specific geometry.

\subsection{\label{sec:theory_zt}Zero temperature}

At zero temperature, we study a system governed by the energy functional

\begin{align}
 &E [\psi] = \int d{\bf r} \psi^{*}({\bf r}) \left[ -\frac{\hbar^2 \nabla^2}{2 m} + U({\bf r}) \right] \psi({\bf r}) \nonumber \\
 &+ \frac{1}{2} \int \int d{\bf r} d{\bf r'} \abs{\psi({\bf r})}^2 V({\bf r} - {\bf r'}) \abs{\psi({\bf r'})}^2 \nonumber \\
 &+ \int d{\bf r} \epsilon_{\rm BMF} \left( \abs{\psi({\bf r})}^2 \right)
 \label{energy}
\end{align}
where $U({\bf r}) = \frac{1}{2} m \omega^2 z^2 + U_{\rm box}(x,y)$ with $U_{\rm box}(x,y)$ a box trap potential. $V({\bf r} - {\bf r'})$ is the contact plus dipole-dipole interaction, i.e.
\begin{align}
 V({\bf r} - {\bf r'}) = \frac{4 \pi \hbar^2 a}{m} \delta\left( {\bf r} - {\bf r'} \right) + \frac{C_{\rm dd}}{4 \pi} \frac{ \left( 1 - 3 \cos^2 \theta_{\alpha} \right) }{\abs{{\bf r} - {\bf r'}}^3} \ , \label{pseudopot}
\end{align}
with $a$ the s-wave scattering length, $C_{\rm dd} = \mu_0 \mu^2$ for magnetic dipoles and $\theta_{\alpha}$ the angle between the vector ${\bf r} - {\bf r'}$ and the polarization axis of the dipoles. In the calculations, we change this direction within the $x$-$z$ plane, and define $\alpha$ as the angle between the polarization direction and the $z$-axis, as in Ref.~\cite{baena2025:pra}. We restrict ourselves to the regime where the collisions between particles can be treated as a three-dimensional process, meaning that $l > a$ with $l = \sqrt{\frac{\hbar}{m \omega}}$, which is necessary for the validity of Eq.~\ref{pseudopot}. We define characteristic energy and length scales given by $r_0 = 12 \pi a_{\rm dd}$ and $E_0 = \hbar^2 / (m r_0^2)$ and present our results in these units. Here, $a_{\rm dd} = m C_{\rm dd}/(12 \pi \hbar^2)$ is the dipole length. For the experimentally relevant cases of $^{164}$Dy and $^{166}$Er, the values of $a_{\rm dd}$ and $r_0$ are $a_{\rm dd}^{\rm Dy} = 130.8 a_0$, $r_0^{\rm Dy} = 0.26$ $\mu$m and $a_{\rm dd}^{\rm Er} = 65.5 a_0$, $r_0^{\rm Er} = 0.13$ $\mu$m, where $a_0$ is the Bohr radius.

The quantity $\epsilon_{\rm BMF} \left( n = \abs{\psi({\bf r})}^2 \right)$ is an energy functional accounting for beyond mean-field (BMF) effects. Performing a functional minimization of Eq.~\ref{energy} with respect to $\psi^{*}({\bf r})$ with the constrain $N = \int d{\bf r} \abs{\psi({\bf r})}^2$ we obtain the zero temperature, extended Gross-Pitaevskii equation (eGPE), which is given by
\begin{align}
 &\mu \psi({\bf r}) = \left[ -\frac{\hbar^2 \nabla^2}{2 m} + U({\bf r}) + \int d{\bf r'} V({\bf r} - {\bf r'}) \abs{\psi({\bf r'})}^2 \right. \nonumber \\
 &\left. + \eval{\pdv{\epsilon_{\rm BMF}}{n}}_{n=\abs{\psi({\bf r})}^2} \right] \psi({\bf r}) \ .
 \label{eGPE}
\end{align}
The BMF energy functional of Eq.~\ref{energy} is computed following the prescription of Refs.~\cite{Zin_2021,baena2025:pra} to account for the discretized nature of the excitations along the $z$-axis. It can be written as
\begin{equation}
 \eval{\pdv{\epsilon_{\rm BMF}}{n}}_{n=\abs{\psi({\bf r})}^2} = \eval{C n^{3/2}}_{n=\abs{\psi({\bf r})}^2} = C \abs{\psi({\bf r})}^{3}
\end{equation}
with $C$ a fitting constant. In the regime of densities considered in this work, the resulting BMF correction to the chemical potential, i.e. $\mu_{\rm BMF} = \eval{\pdv{\epsilon_{\rm BMF}}{n}}_{n=\abs{\psi({\bf r})}^2}$ differs in less than $10 \%$ from the usual beyond mean-field correction for a dipolar system in three-dimensions~\cite{Lima:2011eq, pelster12}, i.e.
\begin{align}
 \mu_{\rm BMF, 3D} =\eval{ \frac{32 g \sqrt{a^3}}{3 \sqrt{\pi}} \mathcal{Q}_{5}\left( \frac{a_{\rm dd}}{a} \right) n^{3/2} }_{n=\abs{\psi({\bf r})}^2} \ .
\end{align}
where $g = 4 \pi \hbar^2 a/m$ and and
$Q_5(\varepsilon_{\mathrm{dd}}) = \dfrac{1}{2} \displaystyle \int_{0}^{\pi}
\mathrm{d} \alpha \sin\alpha \left[1 + \varepsilon_{\rm dd} (3 \cos^2 \alpha -
1)\right]^{5/2}$.

\subsection{\label{sec:theory_ft}Finite temperature}

In order to account for the effect of finite temperature, we need to obtain the correction by thermal fluctuations to Eq.~\ref{eGPE}. To do so, we make use of the local density approximation (LDA). We consider initially a system of tilted dipoles infinite and homogeneous along the $x$-$y$ plane and harmonically trapped along the $z$-axis (the effect of the trapping in the $x$-$y$ plane is considered later). The Bogoliubov-de Gennes equations are:
\begin{widetext}
\begin{align}
&E_j u_j({\bf r}) = \left( -\frac{\hbar^2}{2m} \nabla^2 + U_{\rm osc}(z) - \mu \right) u_j({\bf r}) + u_j({\bf r}) \int d{\bf r'} V({\bf r}-{\bf r'}) \abs{\psi_0({\bf r'})}^2  \nonumber \\
&+ \psi_0({\bf r}) \int d{\bf r'} V({\bf r}-{\bf r'}) \psi_0({\bf r'}) u_j({\bf r'}) - \psi_0({\bf r}) \int d{\bf r'} V({\bf r}-{\bf r'}) \psi_0({\bf r'}) v_j({\bf r'}) \label{bdg1}
\\
&-E_j v_j({\bf r}) = \left( -\frac{\hbar^2}{2m} \nabla^2 + U_{\rm osc}(z) - \mu \right) v_j({\bf r}) + v_j({\bf r}) \int d{\bf r'} V({\bf r}-{\bf r'}) \abs{\psi_0({\bf r'})}^2 \nonumber \\
&+ \psi_0({\bf r}) \int d{\bf r'} V({\bf r}-{\bf r'}) \psi_0({\bf r'}) v_j({\bf r'}) - \psi_0({\bf r}) \int d{\bf r'} V({\bf r}-{\bf r'}) \psi_0({\bf r'}) u_j({\bf r'})  \label{bdg2} \ ,
\end{align}
\end{widetext}
where $U_{\rm osc}(z) = \frac{1}{2} m \omega^2 z^2$. $\psi_0({\bf r})$ and $\mu$ are obtained by solving Eq. ~\ref{eGPE} without both the BMF term and the box trap. We can write the condensate wave function as $\psi_0({\bf r}) = \sqrt{n_{0}} \psi_0(z)$, with $n_{0}$ the 2D condensate density. Thus, the condensate wave function is normalized to the condensate atom number, $N_0 = \int d{\bf r} \abs{\psi_0({\bf r})}^2$ with $\int dz \abs{\psi_0(z)}^2 = 1$. Since the system is infinite in the $x$-$y$ plane, the momentum ${\bf k_{\perp}} = (k_x,k_y)$ is a good quantum number for the longitudinal excitations, Hence, we expand the Bogoliubov amplitudes as
\begin{align}
 u_j({\bf r}) &= u_j({\bf k_{\perp}},z) e^{i {\bf k_{\perp}} {\bf r_{\perp}}} \label{ubg} \\
 v_j({\bf r}) &= v_j({\bf k_{\perp}},z) e^{i {\bf k_{\perp}} {\bf r_{\perp}}} \label{vbg}
\end{align}
where the usual normalization condition $\int dz \left[ \abs{u_j({\bf k}_{\perp},z)}^2 - \abs{v_j({\bf k}_{\perp},z)}^2 \right] = 1$ is satisfied. Inserting Eqs.~\ref{ubg} and~\ref{vbg} into Eqs.~\ref{bdg1} and~\ref{bdg2}, multiplying both sides of the resulting equations by $e^{-i {\bf k_{\perp}}}$ and integrating over $x$ and $y$ leads to two coupled integro-differential equations for the amplitudes $u_j({\bf k_{\perp}},z)$ and $v_j({\bf k_{\perp}},z)$ and the excitation energies $E_j$
\begin{widetext}
\begin{align}
E_j u_j({\bf k_{\perp}},z) &= \left( \frac{\hbar^2 {\bf k}_{\perp}^2}{2m} -\frac{\hbar^2}{2m} \dv[2]{}{z} + U_{\rm osc}(z) - \mu \right) u_j({\bf k_{\perp}},z) +
n_0 \abs{\psi_0(z)}^2 V(0,0) u_j({\bf k_{\perp}},z) \nonumber \\
& +
n_0\int dz' M({\bf k_{\perp}},z,z') u_j({\bf k_{\perp}},z') -n_0\int dz' M({\bf k_{\perp}},z,z') v_j({\bf k_{\perp}},z')
\label{bdg1_full}
\\
-E_j v_j({\bf k_{\perp}},z) &= \left( \frac{\hbar^2 {\bf k}_{\perp}^2}{2m} -\frac{\hbar^2}{2m} \dv[2]{}{z} + U_{\rm osc}(z) - \mu \right) v_j({\bf k_{\perp}},z) +
n_0 \abs{\psi_0(z)}^2 V(0,0) v_j({\bf k_{\perp}},z) \nonumber \\
&+
n_0\int dz' M({\bf k_{\perp}},z,z') v_j({\bf k_{\perp}},z') - n_0\int dz' M({\bf k_{\perp}},z,z') u_j({\bf k_{\perp}},z')
\label{bdg2_full}
\\
M({\bf k_{\perp}},z,z') &= \frac{1}{2 \pi} \int dk_z V({\bf k}) \psi_0(z) e^{i k_z z} \psi_0(z') e^{-i k_z z'} = \psi_0(z) \psi_0(z') V({\bf k_{\perp}},z-z')
\\
V({\bf k}) &= \int d{\bf k} V({\bf r}) e^{-i {\bf k}{\bf r}} = \frac{4 \pi \hbar^2}{m} \left[ a + a_{\rm dd} \left( 3 \frac{(k_x \sin \alpha + k_z \cos \alpha)^2}{k^2} - 1 \right) \right]
\label{V_FT}
\\
V(0,0) &= \eval{ V({\bf k}) }_{k_x = k_y = 0} = \frac{4 \pi \hbar^2}{m} \left[ a + a_{\rm dd} \left( 3 \cos^2 \alpha - 1 \right) \right]
\label{V_FT_0}
\end{align}
\end{widetext}
Eqs.~\ref{bdg1_full}~\ref{bdg2_full} can, in principle, be solved numerically, although this has a considerable computational cost. In order to significantly reduce it, we make the following approximation for the amplitudes $u_j({\bf k_{\perp}},z)$, $v_j({\bf k_{\perp}},z)$ taking advantage of the quasi-2D condition of the system induced by the strong trapping along the $z$ axis.
\begin{align}
 u_j({\bf k_{\perp}},z) &\simeq u_j({\bf k_{\perp}}) \psi_j(z)
 \label{ubg_approx}
 \\
 v_j({\bf k_{\perp}},z) &\simeq v_j({\bf k_{\perp}}) \psi_j(z)
 \label{vbg_approx}
\end{align}
where $\psi_j(z)$ are the eigenstates of the Gross-Pitaevskii equation satisfied by the condensate wave function, i.e.
\begin{align}
 &\mu \psi_j(z) = \left[ -\frac{\hbar^2}{2 m} \dv[2]{}{z} + U_{\rm osc}(z) \right. \nonumber \\
 &\left. + n_0\int d{\bf r'} V({\bf r} - {\bf r'}) \abs{\psi_0(z')}^2 \right] \psi_j(z) \ .
 \label{GPE}
\end{align}
which we normalize as $\int dz \abs{\psi_j(z)}^2 = 1$.

We insert Eqs.~\ref{ubg_approx}~\ref{vbg_approx} into Eqs.~\ref{bdg1_full}~\ref{bdg2_full}, multiply both sides by $\psi_j(z)$ and integrate over $z$ to reach the equations
\begin{widetext}
\begin{align}
E_j u_j({\bf k_{\perp}}) &= \left( \frac{\hbar^2 {\bf k}_{\perp}^2 }{2m} + \Delta E_j \right) u_j({\bf k_{\perp}})
+ n_0 V_j({\bf k}_{\perp}) u_j({\bf k_{\perp}})
- n_0 V_j({\bf k}_{\perp}) v_j({\bf k_{\perp}})
\label{bdg1_approx}
\\
-E_j v_j({\bf k_{\perp}}) &= \left( \frac{\hbar^2 {\bf k}_{\perp}^2 }{2m} + \Delta E_j \right) v_j({\bf k_{\perp}})
+ n_0 V_j({\bf k}_{\perp}) v_j({\bf k_{\perp}})
- n_0 V_j({\bf k}_{\perp}) u_j({\bf k_{\perp}})
\label{bdg2_approx}
\\
V_j({\bf k}_{\perp}) &= \int dz dz' \psi_j(z) \psi_0(z) \psi_j(z') \psi_0(z') V({\bf k}_{\perp},z-z')
\label{V_FT_j}
\\
\Delta E_j &= \left[ \int dz \left( -\frac{\hbar^2}{2 m} \psi_j(z) \dv[2]{\psi_j(z)}{z} + U_{\rm osc}(z) \abs{\psi_j(z)}^2 + n_0 V(0,0) \abs{\psi_j(z)}^2 \abs{\psi_0(z)}^2 \right) - \mu \right] \ .
\end{align}
\end{widetext}
Note that, since $\psi_0$ is an eigenstate of Eq.~\ref{GPE} with eigenvalue $\mu$, $\Delta E_0 = 0$. Eqs.~\ref{bdg1_approx} and~\ref{bdg2_approx} have the same structure as the Bogoliubov de Gennes equations for an homogeneous 3D system, with the addition of the diagonal term $\Delta E_j$. The excitation spectrum and the Bogoliubov amplitudes thus satisfy~\cite{oktel19}
\begin{widetext}
\begin{align}
E_j({\bf k}_{\perp}) &= \sqrt{ \left( \frac{\hbar^2 {\bf k}_{\perp}^2 }{2m} + \Delta E_j \right) \left( \frac{\hbar^2 {\bf k}_{\perp}^2 }{2m} + \Delta E_j + 2 n_0 V_j({\bf k}_{\perp}) \right) } \label{exc_spectrum}
\\
\abs{v_j({\bf k}_{\perp})}^2 &= \frac{1}{2 E_j({\bf k}_{\perp})} \left( \frac{\hbar^2 {\bf k}_{\perp}^2 }{2m} + \Delta E_j + n_0 V_j({\bf k}_{\perp}) - E_j({\bf k}_{\perp}) \right)
\label{v_final}
\\
\abs{u_j({\bf k}_{\perp})}^2 &= 1 + \abs{v_j({\bf k}_{\perp})}^2
\label{u_final}
\\
u_j({\bf k}_{\perp}) v^*_j({\bf k}_{\perp}) &= \frac{n_0 V_j({\bf k}_{\perp})}{ 2 E_j({\bf k}_{\perp}) }
\label{uv_final}
\end{align}
\end{widetext}
We show in the Appendix~\ref{sec:app_exc_spectrum} a comparison between the results for the excitation spectrum along the $k_y$ axis obtained from Eqs.~\ref{bdg1_approx}~\ref{bdg2_approx} and the full solution of Eqs.~\ref{bdg1_full}~\ref{bdg2_full}, which was computed in Ref.~\cite{Baillie_2015}. Such comparison shows excellent agreement between both approaches.
Retaining only the contribution of the lowest mode, $j=0$, corresponds to the same shape approximation discussed in Ref.~\cite{Baillie_2015}.

Using Eqs.~\ref{exc_spectrum}-~\ref{uv_final}, we can obtain the temperature dependent contribution to the grand canonical potential energy density, as well as the 2D density of the thermal cloud, which are given by~\cite{Gaul2014:APB,oktel19,baena22}
\begin{align}
 \frac{\Omega_{\rm th}}{S} &= \frac{1}{\beta} \sum_{j=0}^{\infty} \int \frac{ d{\bf k_{\perp}} }{(2 \pi)^2} \ln \left( 1 - e^{-\beta E_j({\bf k}_{\perp})} \right)
 \label{gc_th}
 \\
 n_{\rm th} &= \sum_{j=0}^{\infty} \int \frac{ d{\bf k_{\perp}} }{(2 \pi)^2} \frac{1}{e^{\beta E_j({\bf k}_{\perp})}-1} \left( \abs{u_j({\bf k}_{\perp})}^2 + \abs{v_j({\bf k}_{\perp})}^2 \right)
 \label{n_th}
\end{align}
where $\beta = 1/(k_B T)$ and $k_B$ is the Boltzmann constant.

Thus far, we have considered an infinite system in the $x$-$y$ plane for the calculation of the thermal excitations. However, as mentioned previously in Sec.~\ref{sec:theory_zt}, the thermal density $n_{\rm th}$ diverges for any temperature $T > 0$ under these conditions, which is a consequence of the absence of a BEC in two-dimensions in the thermodynamic limit.  However, the condensate can be restored if one considers a trapping potential in the $x$-$y$ plane. This can be incorporated in our calculation for the thermal excitations by introducing a low momentum cut-off in Eqs.~\ref{gc_th} and~\ref{n_th} given by the trap. As mentioned previously, we consider a box trap, which is given by
\begin{align}
\small
 U_{\rm box}(x,y) =
 \begin{cases}
  0 &\text{ if } x \in [-L_x/2, L_x/2],y \in [-L_y/2, L_y/2]  \\
  \infty &\text{ otherwise }
 \end{cases}
 \label{box_trap}
\end{align}
For a non-interacting gas inside the box trap, the energy spectrum is given by $E(n_x,n_y) = \frac{\hbar^2 \pi^2}{2 m} \left( \frac{n_x^2}{L_x^2} + \frac{n_y^2}{L_y^2} \right)$ for $n_x \geq 1$, $n_y \geq 1$. The chemical potential of the ground state is given by $\mu = E(1,1) = \frac{\hbar^2 \pi^2}{2 m} \left( \frac{1}{L_x^2} + \frac{1}{L_y^2} \right)$, and the number of thermal atoms of the trapped ideal gas is thus given by
\begin{equation}
 N_{\rm th} = \sum_{n_x = 1} \sum_{n_y = 1} \frac{\left( 1 - \delta_{n_x}\delta_{n_y} \right) }{e^{\beta \left( E(n_x,n_y)-\mu \right)} - 1}
\end{equation}

\begin{figure}[t]
\centering
\includegraphics[width=0.85\linewidth]{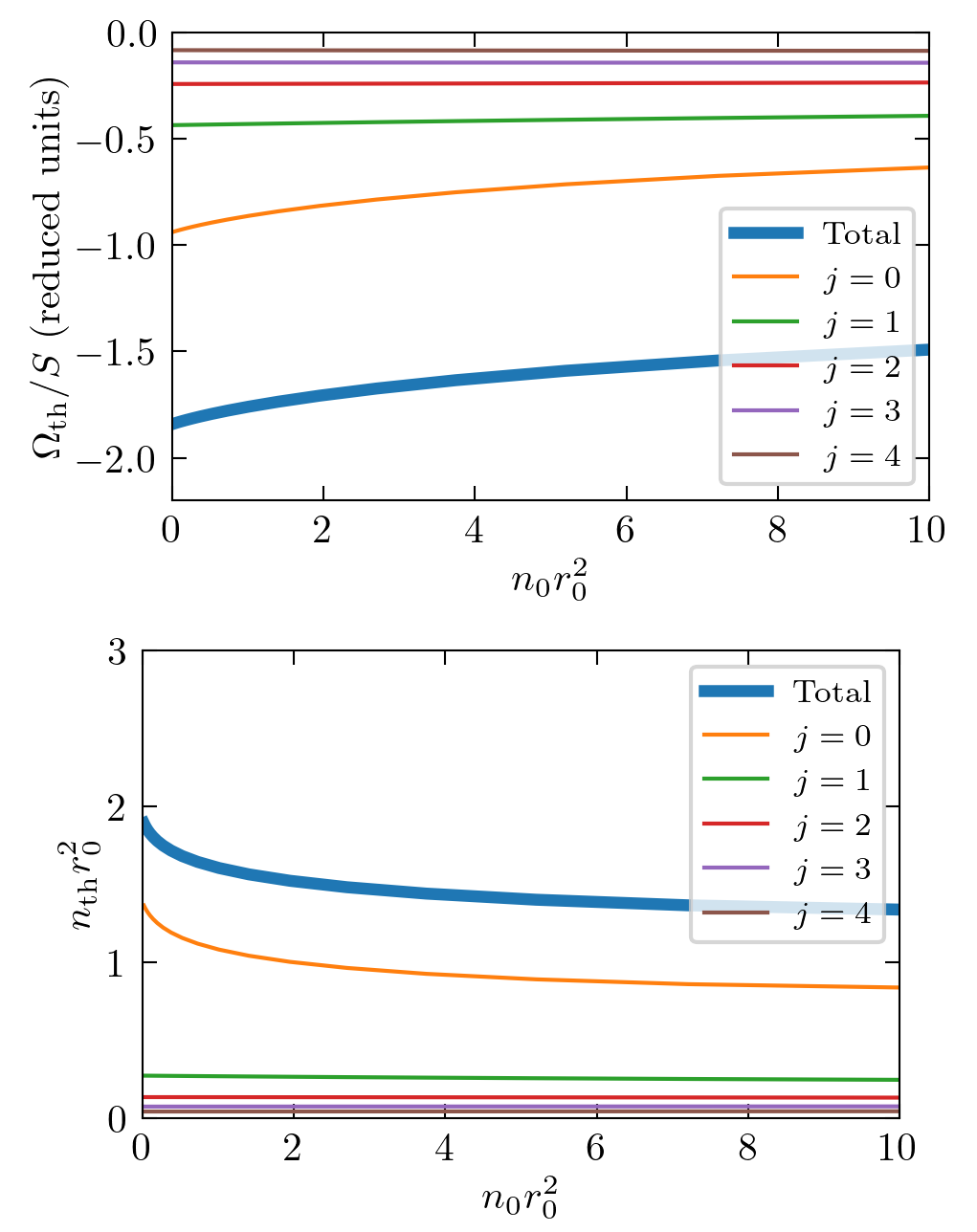}
\caption{Contributions of the different axial excitation modes to the density functionals $\frac{\Omega_{\rm th}}{S} (n_0)$ (2D thermal grand canonical energy density) and $n_{\rm th} (n_0)$ (2D thermal density of depleted atoms), where $n_0$ is the 2D condensate density. The parameters of the calculation are $\hbar \omega/E_0 = 1$, $a/a_{\rm dd} = 0.7$, $\alpha = 64.17^o$, $k_B T/E_0 = 2$. The momentum cut-offs in Eqs.~\ref{gc_th_co}~\ref{n_th_co} are computed for a box trap of widths $L_x/r_0 = L_y/r_0 = 41.02$.}
\label{fig1}
\end{figure}

where $\delta_n$ is the Kronecker delta. For the two first terms of the series, ($n_x = 1$, $n_y = 2$) and ($n_x = 2$, $n_y = 1$), the quantity $E-\mu$ yields $E(2,1)-\mu = \frac{3 \hbar^2 \pi^2}{2 m L_x^2}$ and $E(1,2)-\mu = \frac{3 \hbar^2 \pi^2}{2 m L_y^2}$, respectively. Thus, to account for the finite size of the box trap in Eqs.~\ref{gc_th} and~\ref{n_th}, we can adopt momentum cut-offs in the $x$ and $y$ axes given by $(k_x^{\rm c.o.},k_y^{\rm c.o.}) = \sqrt{3} \pi \left( \frac{1}{L_x}, \frac{1}{L_y} \right)$. Under this consideration, we rewrite the integrals of Eqs.~\ref{gc_th} and~\ref{n_th} as
\begin{widetext}
\begin{align}
 \frac{\Omega_{\rm th}}{S} &= \frac{1}{\beta} \sum_{j=0}^{\infty} \int_{\abs{k_x}>\frac{\sqrt{3}\pi}{L_x},\abs{k_y}>\frac{\sqrt{3}\pi}{L_y}} \frac{ d{\bf {\bf k}_{\perp}} }{(2 \pi)^2} \ln \left( 1 - e^{-\beta E_j({\bf k}_{\perp})} \right)
 + \frac{1}{\beta} \sum_{j=0}^{\infty} \int_{\abs{k_x}=0,\abs{k_y}>\frac{\sqrt{3}\pi}{L_y}} \frac{ dk_y }{(2 \pi L_x)} \ln \left( 1 - e^{-\beta E_j(k_y)} \right) \nonumber \\
 &+ \frac{1}{\beta} \sum_{j=0}^{\infty} \int_{\abs{k_x}>\frac{\sqrt{3}\pi}{L_x},\abs{k_y}=0} \frac{ dk_x }{(2 \pi L_y)} \ln \left( 1 - e^{-\beta E_j(k_x)} \right)
 + \frac{1}{\beta} \sum_{j>0}^{\infty} \frac{1}{L_x L_y} \eval{ \left\{ \ln \left( 1 - e^{-\beta E_j({\bf k}_{\perp})} \right) \right\} }_{{\bf k}_{\perp}=0}
 \label{gc_th_co}
 \\
 n_{\rm th} &= \sum_{j=0}^{\infty} \int_{\abs{k_x}>\frac{\sqrt{3}\pi}{L_x},\abs{k_y}>\frac{\sqrt{3}\pi}{L_y}} \frac{ d{\bf k_{\perp}} }{(2 \pi)^2} \frac{1}{e^{\beta E_j({\bf k}_{\perp})}-1} \left( \abs{u_j({\bf k}_{\perp})}^2 + \abs{v_j({\bf k}_{\perp})}^2 \right)\nonumber \\
 & + \sum_{j=0}^{\infty} \int_{\abs{k_x}=0,\abs{k_y}>\frac{\sqrt{3}\pi}{L_y}} \frac{ dk_y }{(2 \pi L_x)} \frac{1}{e^{\beta E_j(k_y)}-1} \left( \abs{u_j(k_y)}^2 + \abs{v_j(k_y)}^2 \right) \nonumber \\
 &+ \sum_{j=0}^{\infty} \int_{\abs{k_x}>\frac{\sqrt{3}\pi}{L_x},\abs{k_y}=0} \frac{dk_x}{(2 \pi L_y)} \frac{1}{e^{\beta E_j(k_x)}-1} \left( \abs{u_j(k_x)}^2 + \abs{v_j(k_x)}^2 \right) \nonumber \\
 &+ \sum_{j>0}^{\infty} \frac{1}{L_x L_y} \eval{ \left\{  \frac{1}{e^{\beta E_j({\bf k}_{\perp})}-1} \left( \abs{u_j({\bf k}_{\perp})}^2 + \abs{v_j({\bf k}_{\perp})}^2 \right) \right\}  }_{{\bf k}_{\perp}=0}
 \label{n_th_co}
\end{align}
\end{widetext}
In practice, we retain a finite number of modes when computing the integrals in Eqs.~\ref{gc_th_co} and~\ref{n_th_co}. We show in Fig.~\ref{fig1} an example of the contributions with a different value of $j$ to $\frac{\Omega_{\rm th}}{S}$ and $n_{\rm th}$ as a function of the 2D condensate density $n_0$. We find that it is sufficient to retain modes up to $j=4$.

Having obtained the correction by thermal fluctuations to the grand canonical potential, we can now derive a finite temperature equation for a trapped dipolar system in the quasi-2D limit in the style of Eq.~\ref{eGPE} with the inclusion of a temperature-dependent correction. This is the temperature-dependent extended Gross-Pitaevskii equation (TeGPE), already employed in Refs.~\cite{oktel19,baena22,baena24,He2024:PRR} to study 3D systems. In the limit of a large condensate fraction ($n \simeq n_0$, with $n$ the total 2D density and $n_0$ the condensed 2D density), the full grand canonical potential of the system is given by~\cite{baena22}
\begin{align}
 \Omega = E_{\rm MF} + E_{\rm LHY} - \mu N + \Omega_{\rm th} \ ,
 \label{gc_full}
\end{align}
where $E_{\rm MF}$ is the mean-field energy, $E_{\rm LHY}$ is the energy correction by quantum fluctuations and $\Omega_{\rm th}$ is given by Eq.~\ref{gc_th_co}. We can make use of the LDA and replace the 3D density by the square of the wave function ($n_{\rm 3D} \rightarrow \abs{\psi({\bf r})}^2$) and the 2D density by its integral over $z$ ($n \rightarrow \int dz \abs{\psi({\bf r})}^2$). This yields the following expressions for each term~\cite{baena22}
\begin{widetext}
\begin{align}
 E_{\rm MF} &= \int d{\bf r} \left\{ -\psi^*({\bf r}) \frac{\hbar^2}{2m} \nabla^2\psi({\bf r}) + U({\bf r}) \abs{\psi({\bf r})}^2 + \frac{1}{2} \int d{\bf r'} V({\bf r}-{\bf r'}) \abs{\psi({\bf r'})}^2 \abs{\psi({\bf r})}^2 \right\}
 \label{E_MF}
 \\
 E_{\rm LHY} &= \frac{2C}{5} \int d{\bf r} \abs{\psi({\bf r})}^5
 \label{E_LHY}
 \\
 N &\simeq \int d{\bf r} \abs{\psi({\bf r})}^2 \ .
 \label{N_MF}
 \\
 \Omega_{\rm th} &= \int d{\bf r_{\perp}} \eval{ \frac{\Omega_{\rm th}}{S}(n_0) }_{n_0 = \int dz \abs{\psi({\bf r})}^2}
 \label{OMEGA_TH}
\end{align}
\end{widetext}
The constant $C$ of Eq.~\ref{E_LHY} is obtained by fitting the LHY correction to the chemical potential as described in Ref.~\cite{baena2025:pra}. The TeGPE can be obtained by inserting Eqs-~\ref{E_MF}-~\ref{OMEGA_TH} into Eq.~\ref{gc_full} and performing a functional minimization with respect to $\psi^*({\bf r})$, i.e. $\frac{\delta \Omega}{\delta \psi^*} = 0$. This is equivalent to minimizing the free energy $F = \Omega + \mu N$ while fixing the particle number, $N \simeq \int d{\bf r} \abs{\psi({\bf r})}^2$. The functional variation of the terms of Eqs.~\ref{E_MF}-~\ref{N_MF} is performed as in Ref.~\cite{baena22}, while that of Eq.~\ref{OMEGA_TH} results into
\begin{align}
 &\delta \Omega_{\rm th} = \int d{\bf r_{\perp}} \eval{ \pdv{ \left( \Omega_{\rm th}/S \right)}{n_0} }_{n_0 = \int dz \abs{\psi({\bf r})}^2} \frac{\delta n_0}{\delta \psi^*} \delta \psi^* \nonumber \\
 &= \int d{\bf r_{\perp}} dz \eval{ \pdv{ \left( \Omega_{\rm th}/S \right)}{n_0} }_{n_0 = \int dz \abs{\psi({\bf r})}^2} \psi({\bf r}) \delta \psi^* \nonumber \\
 &= \int d{\bf r} \eval{ \pdv{ \left( \Omega_{\rm th}/S \right)}{n_0} }_{n_0 = \int dz \abs{\psi({\bf r})}^2} \psi({\bf r}) \delta \psi^*
\end{align}
This leads to the TeGPE, which is given by
\begin{align}
  &\mu \psi({\bf r}) = \left[ -\frac{\hbar^2 \nabla^2}{2 m} + U({\bf r}) + \int d{\bf r'} V({\bf r} - {\bf r'}) \abs{\psi({\bf r'})}^2 \right. \nonumber \\
 &\left. + C \abs{\psi({\bf r})}^3 + \eval{ \pdv{ \left( \Omega_{\rm th}/S \right)}{n_0} }_{n_0 = \int dz \abs{\psi({\bf r})}^2} \right] \psi({\bf r}) \ ,
 \label{TeGPE}
\end{align}
where the trapping potential is given by $U({\bf r}) = U_{\rm box}({\bf r_{\perp}}) + \frac{1}{2} m \omega^2 z^2$, with $U_{\rm box}({\bf r_{\perp}})$ given in Eq.~\ref{box_trap}. Eq.~\ref{TeGPE} can be solved to obtain the condensate wave function of the system at finite temperature. From the condensate wave function, the 2D density of thermal atoms can be obtained via the expression
\begin{equation}
 n_{\rm th}({\bf r_{\perp}}) = \eval{n_{\rm th} (n_0)}_{n_0 = \int dz \abs{\psi({\bf r})}^2}
 \label{th_density}
\end{equation}
where $n_{\rm th}$ is given in Eq.~\ref{n_th_co} and depends on the 2D condensed density $n_0$. Outside of the box trap, we have $n_{\rm th}({\bf r_{\perp}}) = 0$. Because we have explicitly accounted for the trapping potential along the $z$-axis in the calculation of the thermal density functional, this results into a convergent thermal density.

In practice, to implement the thermal correction of Eq.~\ref{TeGPE} and obtain the thermal density of Eq.~\ref{th_density}, we compute the integrals in Eqs.~\ref{gc_th_co} and~\ref{n_th_co} for several values of the 2D condensed density $n_0$ and fit the results to the following empirical functionals:
\begin{align}
 \frac{\Omega_{\rm th}}{S}(n_0) \simeq -a_1 -a_2 \exp(-a_3 n_0^{a_4})
 \label{fit_1}
 \\
 n_{\rm th}(n_0) \simeq b_1 + b_2 \exp(-b_3 n_0^{b_4})
 \label{fit_2}
\end{align}
with $a_1$, $a_2$, $a_3$, $a_4$, $b_1$, $b_2$, $b_3$ and $b_4$ fitting parameters. Importantly, for sufficiently large values of the tilting angle $\alpha$, the excitation spectrum in Eq.~\ref{exc_spectrum} becomes imaginary at a finite momentum $k_y \neq 0$ for $j=0$, $k_x = 0$ due to the usual roton softening present in confined dipolar systems~\cite{santos2003:prl}. This implies that the results of Eqs.~\ref{gc_th_co}~\ref{n_th_co} acquire an unphysical imaginary part that increases with the 2D condensate density $n_0$. Moreover, the excitation spectrum also suffers a long wavelength instability at tilting angles such that $V(0,0)<0$ (see Eq.~\ref{V_FT_0}). Thus, for the calculation of the excitations, we are restricted to tilting angles and condensate densities where the imaginary part of both $\frac{\Omega_{\rm th}}{S}$ and $n_{\rm th}$ are equal to zero. At the same time, we want to lie as close as possible to the relevant values of tilting angles ($\alpha > 68^o$) experimental conditions. Therefore, for all the cases considered in this work, the excitations and the subsequent thermal correction of Eq.~\ref{TeGPE} are computed for $\alpha=64.17^o$ while the parameters of the functionals in Eq.~\ref{fit_1} and~\ref{fit_2} are obtained by fitting the result of Eqs.~\ref{gc_th_co}~\ref{n_th_co} in the density range $n_0 r_0^2 \in [0,10]$. These parameters serve as an input to the TeGPE. We have checked that our results remain essentially unchanged if we compute the excitations at tilting angles $\alpha=60.16^o$ and $\alpha=68.75^o$ and adjust the range of densities for the fit in order to avoid imaginary contributions (see Appendix~\ref{sec:app_f_t}).

\section{\label{sec:results}Results}

\begin{figure*}[t]
\centering
\includegraphics[width=\linewidth]{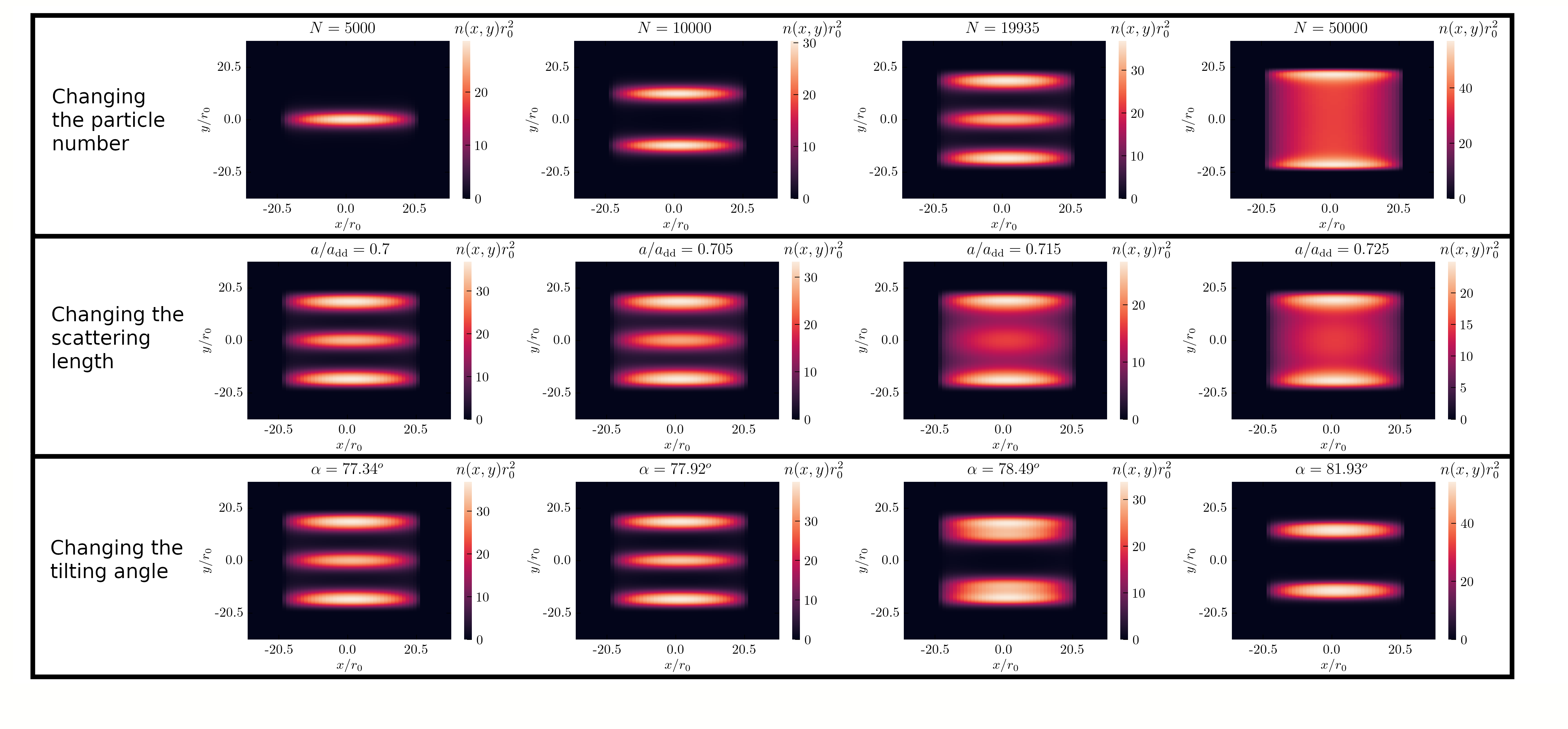}
\caption{Integrated 2D density distribution ($n = \int dz \abs{\psi({\bf r})}^2$) of the dipolar system in a box trap at zero temperature. In each row, one parameter is varied while the rest are fixed to the values $a/a_{\rm dd} = 0.7$, $\alpha = 77.35^o$, $N=19935$. The box trap has widths $L_x/r_0 = L_y/r_0 = 41.02$ while the axial harmonic trap is set to $\hbar \omega /E_0 = 1$.}
\label{fig2}
\end{figure*}

In the following, we present the results for the study of the dipolar system subjected to the trapping potential $U({\bf r}) = U_{\rm box}({\bf r_{\perp}}) + \frac{1}{2} m \omega^2 z^2$, with $U_{\rm box}({\bf r_{\perp}})$ given in Eq.~\ref{box_trap}, both at zero as well as finite temperature. The trapping frequency is fixed to $\hbar \omega / E_0 = 1$ in all cases, which corresponds to $\omega = 2\pi \times 906$ Hz for $^{164}$Dy atoms and $\omega = 2\pi \times 3569$ Hz for $^{166}$Er atoms. In this Section, we denote by $N_0$ the number of atoms in the condensate and by $N$ the total number of atoms. At zero temperature, $N_0 \simeq N$ since our system is dilute enough to neglect the quantum depletion. The maximum gas parameter in our system is of order $x_{\rm max} = n_{\rm 3D, peak} a^3 \sim 10^{-4}$, while the tilting angles considered in this work lay significantly far from the limit $\alpha = 0$ where the quantum depletion diverges even at zero temperature~\cite{fischer_06_PRA,fedorov_2014_PRA}. At finite temperature, on the other hand, $N_0 \neq N$ due to the thermal effects. Eqs.~\ref{eGPE} and~\ref{TeGPE} are numerically solved via imaginary time propagation to obtain the equilibrium condensate wave function at zero and finite temperature, respectively. We discuss the technical aspects of the numerical simulations in the Appendix~\ref{sec:app_tech}.

\subsection{\label{sec:zero_T}Zero temperature}

We start by illustrating how the structure of the condensate changes as a function of the main parameters of our system (the particle number $N$, the tilting angle of the dipoles $\alpha$ and the scattering length $a$) for a given box trap potential. We consider a system confined in a box trap of size $L_x/r_0= L_y/r_0 = 41.02$ and a tilting angle of $\alpha = 77.35^o$. According to the phase diagram of Ref.~\cite{baena2025:pra}, an infinite system in the $x$-$y$ plane with the same tilting lies in the striped liquid region at the equilibrium density, given by $n_{\rm 2D, eq} r_0^2 = 11.84$. For the box trap considered, this corresponds to an atom number of $N = 19935$. For this particle number, the finite system also features striped density modulations. We show in Fig.~\ref{fig2}, the change in the 2D integrated density $n = \int dz \abs{\psi({\bf r})}^2$ when either one of $N$, $\alpha$ or $a$ are varied while keeping the rest of the parameters constant. The results show that an increase of the particle number destroys the density modulations. This behaviour is present in dipolar systems confined in different geometries, from tubular~\cite{Blakie_2020} to planar~\cite{zhang19,zhang21} to shell-shaped~\cite{baena23shell}, and is also present for the infinite system in our quasi-2D setting~\cite{baena2025:pra}. However, in the case of infinite planar geometries less tightly trapped and at lower tilting angles than the ones considered in this work, this effect is only present at high 2D densities of the order of $n r_0^2 \sim 200$~\cite{zhang19,zhang21}. Remarkably, in our finite size system, the loss of modulations takes place at a significantly lower value of the 2D density ($n r_0^2 \simeq 20$ for the chosen parameters). Similarly, an increase in the scattering length also eliminates the density modulations, which matches the well-known behaviour of dipolar systems~\cite{Ferlaino:PRX:2019}. On the other hand, an increase in the tilting angle reduces the number of stripes, which is a consequence of the increased attraction of the DDI, which favors bunching.

\begin{figure*}[t]
\centering
\includegraphics[width=\linewidth]{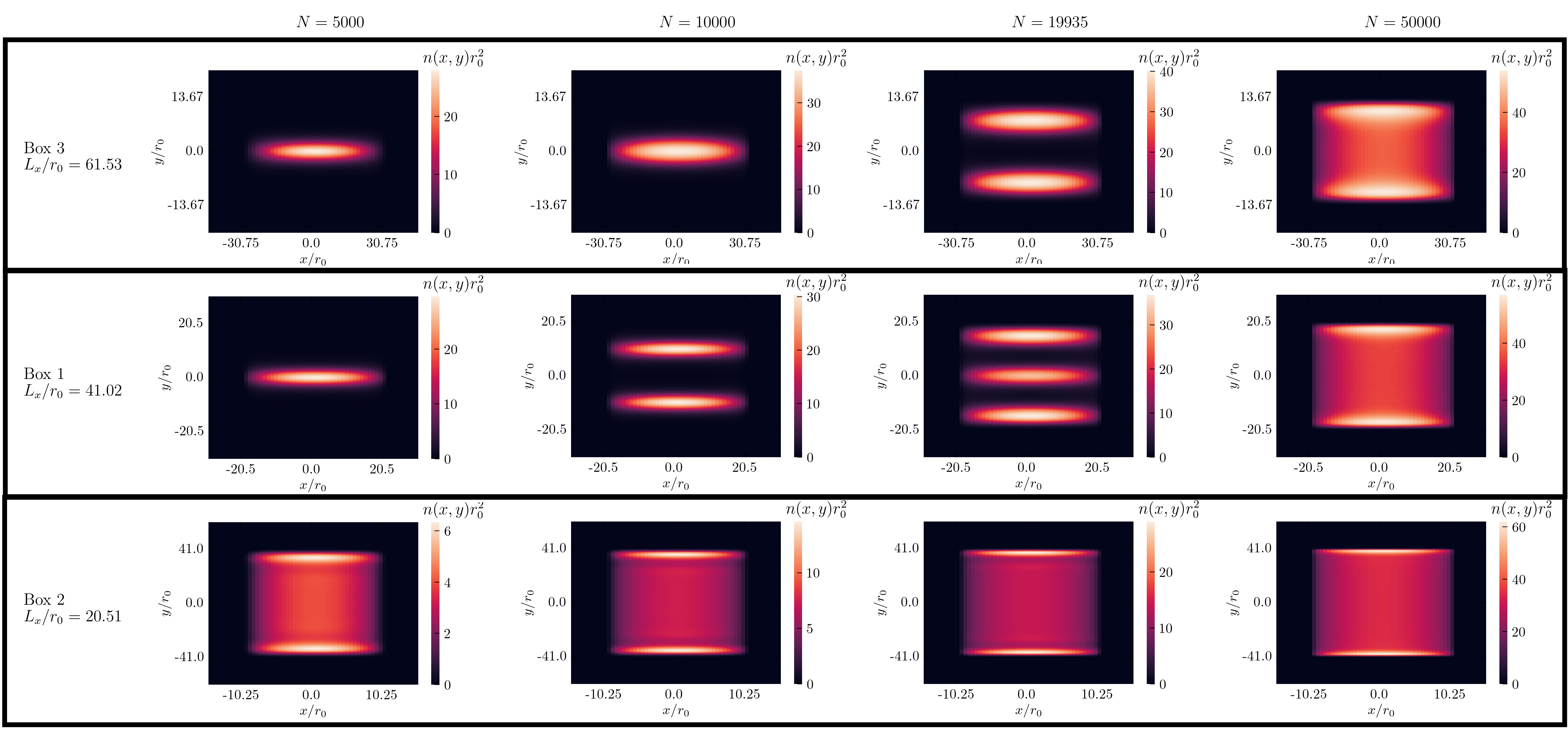}
\caption{Integrated 2D density distribution ($n = \int dz \abs{\psi({\bf r})}^2$) of the dipolar system for box traps with different aspect ratios at zero temperature. Each column and row correspond to a different particle number and box trap, respectively. In each column, the average 2D density is kept fixed. The widths along the $y$-axis of each box trap are given by $L_y/r_0 = 41.02$ for Box 1, $L_y/r_0 = 82.04$ for Box 2 and $L_y/r_0 = 27.36$ for Box 3. The rest of the parameters of the calculations are $a/a_{\rm dd} = 0.7$, $\alpha = 77.35^o$ and $\hbar \omega /E_0 = 1$.}
\label{fig3}
\end{figure*}
\begin{figure}[t]
\centering
\includegraphics[width=\linewidth]{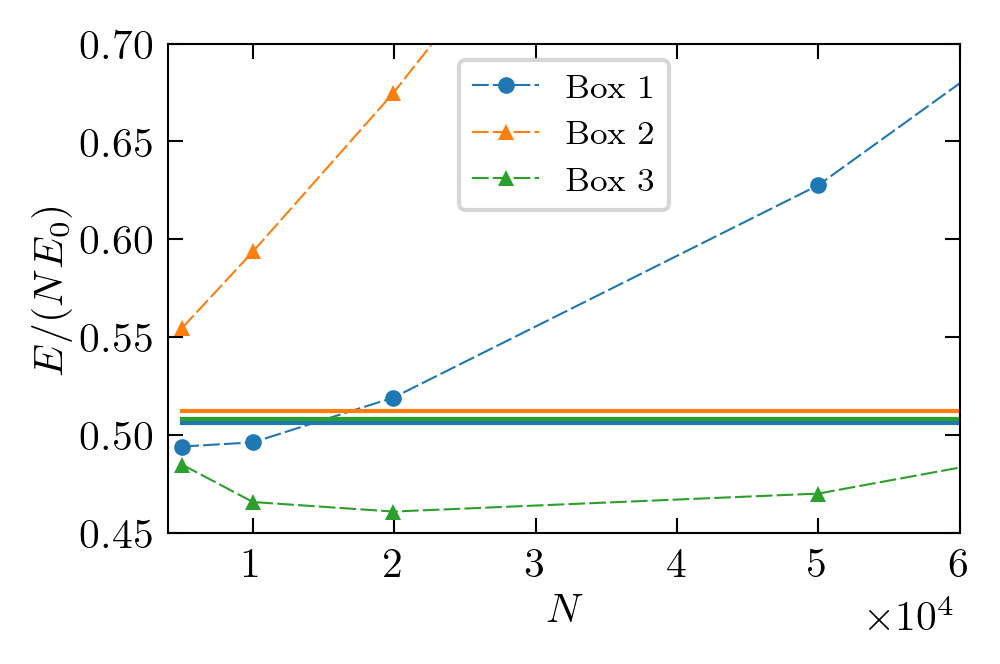}
\caption{Energy per particle as a function of $N$ for the tilted dipolar system in three different box traps. The dashed-dotted lines correspond to the results of the interacting system while the solid lines correspond to the trapped ideal gas. The parameters of the calculations are the same as in Fig.~\ref{fig3}.}
\label{fig4}
\end{figure}

Since we are dealing with an anisotropic system in the $x$-$y$ plane due to the non-zero tilting angle of the dipoles, changing the value of the aspect ratio of the box trap has a relevant effect on the structure of the condensate. To illustrate this, in Fig.~\ref{fig3} we show the results for the 2D density obtained under three different box trap confinements: $L_x/r_0= L_y/r_0 = 41.02$ (or ``Box 1'', same as Fig.~\ref{fig2}), $L_x/r_0 = 20.51$, $L_y/r_0 = 82.04$ (or ``Box 2''), and $L_x/r_0 = 61.53$, $L_y/r_0 = 27.36$ (or ``Box 3''). The two latter boxes correspond to a different aspect ratio, but equal average 2D density. The results show that the different aspect ratios lead to different structures. This is because the size of the box trap along the $x$ axis, $L_x$, controls the bunching of particles to form stripes, which arise in this direction due to the dipoles being polarized in the $x$-$z$ plane. Aspect ratios with larger $L_x$ result into dipoles forming a lower number of longer stripes, while smaller values of $L_x$ frustrate the accumulation of dipoles in a single stripe, first splitting it into several stripes to then destroy modulations. Remarkably, we can see how the atoms occupy all the available space of the box trap with the smallest $L_x$, displaying a behaviour alike to that of a gas, while they do not do so for the largest value of $L_x$, where the system behaves as a liquid. This shows that the aspect ratio of the box trap (or, more precisely, the value of $L_x$) has an important effect on the behaviour of the system. This is reflected in the energy per particle, which we report in Fig.~\ref{fig4}. From the figure, we can see that, for the largest value of $L_x$, the lowest particle numbers display an energy per particle lower than that of the trapped non-interacting gas, $E_{\rm NI}/N = \hbar \omega/2 + \hbar^2 \pi^2/(2 m L_x^2) + \hbar^2 \pi^2/(2 m L_y^2)$. This indicates the system behaves as a liquid. On the other hand, the energies per particle increase significantly for the smallest value of $L_x$ and rise above the non-interacting energy per particle threshold, which results into a gas-like behaviour. These results showcase the importance of finite size effects, which are present in any experimental realization.

\subsection{\label{sec:finite_T}Finite temperature}

\begin{figure}[t]
\centering
\includegraphics[width=0.85\linewidth]{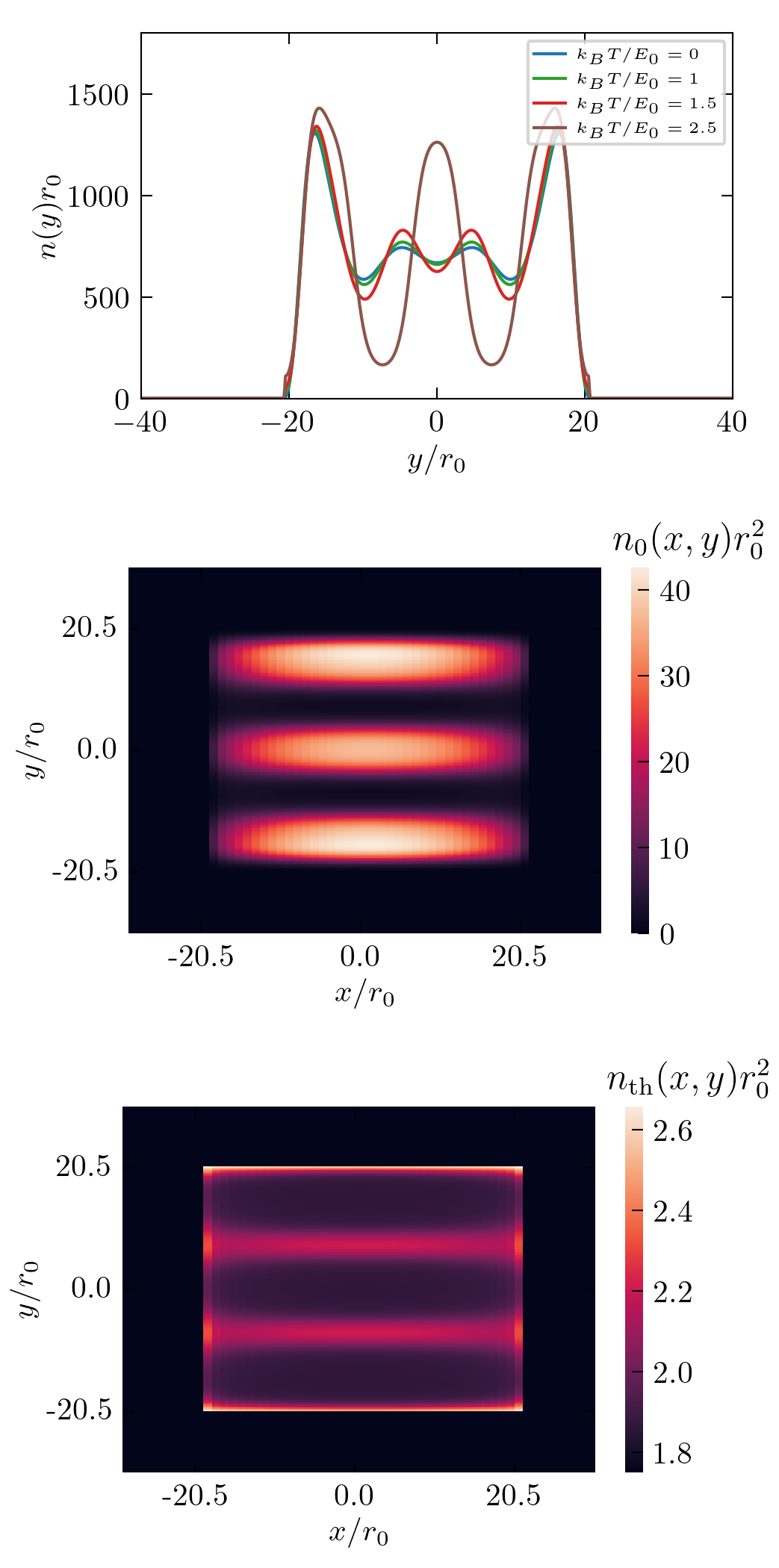}
\caption{Top: total column density $n(y) = \int dx dz \text{ } \abs{\psi({\bf r})}^2 + \int dx \text{ } n_{\rm th}({\bf r}_{\perp})$ for different values of the temperature for the parameters $N=31000$, $a/a_{\rm dd} = 0.7$, $\alpha = 77.35^o$ $L_x/r_0 = L_y/r_0 = 41.02$ and $\hbar \omega /E_0 = 1$. Middle and bottom: condensate and thermal 2D densities, respectively, for $k_B T/E_0 = 2.5$. }
\label{fig5}
\end{figure}

We now move on to consider the inclusion of finite temperature effects, as they are present in any current state of the art experiment. To start, we illustrate the effect of a change of temperature alone for a fixed value of the scattering length $a$, the tilting $\alpha$ and the \textit{total} particle number $N$. It must be emphasized that, unlike in previous works~\cite{oktel19,baena22,baena24,He2024:PRR}, accounting explicitly for the trap along the $z$-axis in the calculation of the thermal excitations of Sec.~\ref{sec:theory_ft} allows us to obtain a convergent population of thermal atoms, enables the study thermal effects in the canonical ensemble and the estimation of the condensate fraction. We consider a system with $a/a_{\rm dd} = 0.7$, $\alpha = 77.35^o$ and $N=31000$ in a box trap with widths $L_x/r_0 = L_y/r_0 = 41.02$ and show, in Fig.~\ref{fig5}, the total atom column density ($n(y) = \int dx dz \text{ } \abs{\psi({\bf r})}^2 + \int dx \text{ } n_{\rm th}({\bf r}_{\perp})$, with $n_{\rm th}({\bf r}_{\perp})$ the 2D thermal density given in Eq.~\ref{th_density} and $\psi({\bf r})$ the solution of Eq.~\ref{TeGPE}) for different values of the temperature between $k_B T / E_0 = 0$ and $k_B T / E_0 = 2.5$. This range of temperatures corresponds to $T \in [0,109.5]$ nK for $^{164}$Dy atoms and $T \in [0, 438]$ nK for $^{166}$Er atoms. In the figure, we also illustrate the 2D condensate and thermal densities, $n_0(x,y) = \int dz \text{ } \abs{\psi({\bf r})}^2$ and $n_{\rm th}(x,y)$ respectively, for $k_B T / E_0 = 2.5$. From the results, we see, remarkably, that the atomic density tends to modulate upon increasing the temperature while leaving the total atom number fixed. This is in qualitative agreement with previous Path Integral Monte Carlo results for a 3D system of dipoles in a tubular geometry~\cite{bombin25:prl}. Previous works employing Bogoliubov theory had shown that this is also the case if the condensate atom number is kept fixed upon increasing the temperature~\cite{baena22,baena24,He2024:PRR}. Remarkably, though, this differs from the behaviour of the infinite 2D system, where dipolar stripes are destroyed by thermal effects~\cite{bombin19:PRA,pengtao2021}. However, and as mentioned previously, in that limit the condensate is completely depleted by thermal fluctuations and the theory presented here is inapplicable. This showcases the importance of the existence of a BEC, which changes the qualitative behavior of the system with respect to temperature. In our case, there are two main effects leading to the observed thermal behaviour. On one hand, and as stated in previous works~\cite{baena22,baena24,He2024:PRR}, the thermal term in Eq.~\ref{TeGPE} acts as a density dependent potential that decreases as the condensate density increases, thus acting as a focusing non-linearity that favors states with high density regions. On the other hand, the zero temperature results of Fig.~\ref{fig2} show that an increase in the number of particles of the condensate tends to destroy modulations. Thus, since we are keeping $N$ constant while rising the temperature, the number of condensate particles drops as the temperature increases, further promoting modulations.
\begin{figure}[t]
\centering
\includegraphics[width=0.85\linewidth]{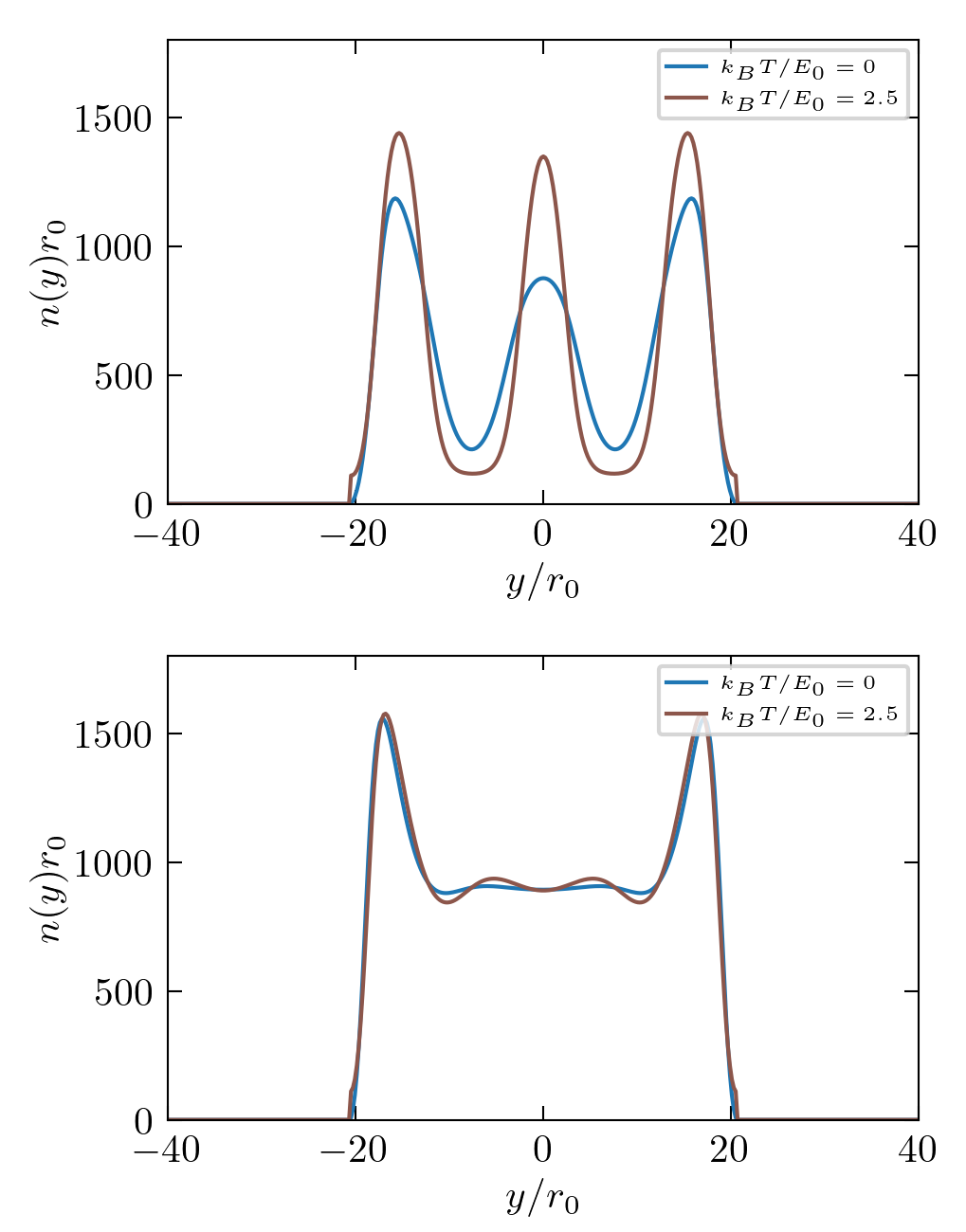}
\caption{Total column density $n(y) = \int dx dz \text{ } \abs{\psi({\bf r})}^2 + \int dx \text{ } n_{\rm th}({\bf r}_{\perp})$ at zero and finite temperature for $N=25000$ (top) and $N=40000$ (bottom). The rest of the parameters are the same as in Fig.~\ref{fig5}. }
\label{fig6}
\end{figure}
Looking at the 2D densities shown in Fig.~\ref{fig5} we can also see that thermal atoms tend to accumulate in the regions of low condensate densities. This behaviour can be understood by looking at Fig.~\ref{fig1}, where we report the density functional of the thermal density $n_{\rm th} (n_0)$. Notice that this density functional decreases as the condensate density increases, meaning that the higher contribution to $n_{\rm th}$ comes from the regions less populated by the condensate. Importantly, since the theory presented in Sec.~\ref{sec:theory_ft} allows to obtain the density of thermal atoms, it could be used for thermometry in state of the art experiments employing a box trap, in order to extract the temperature, number of atoms, or the chemical potential of the experiment.´

\begin{figure*}[t]
\centering
\includegraphics[width=\linewidth]{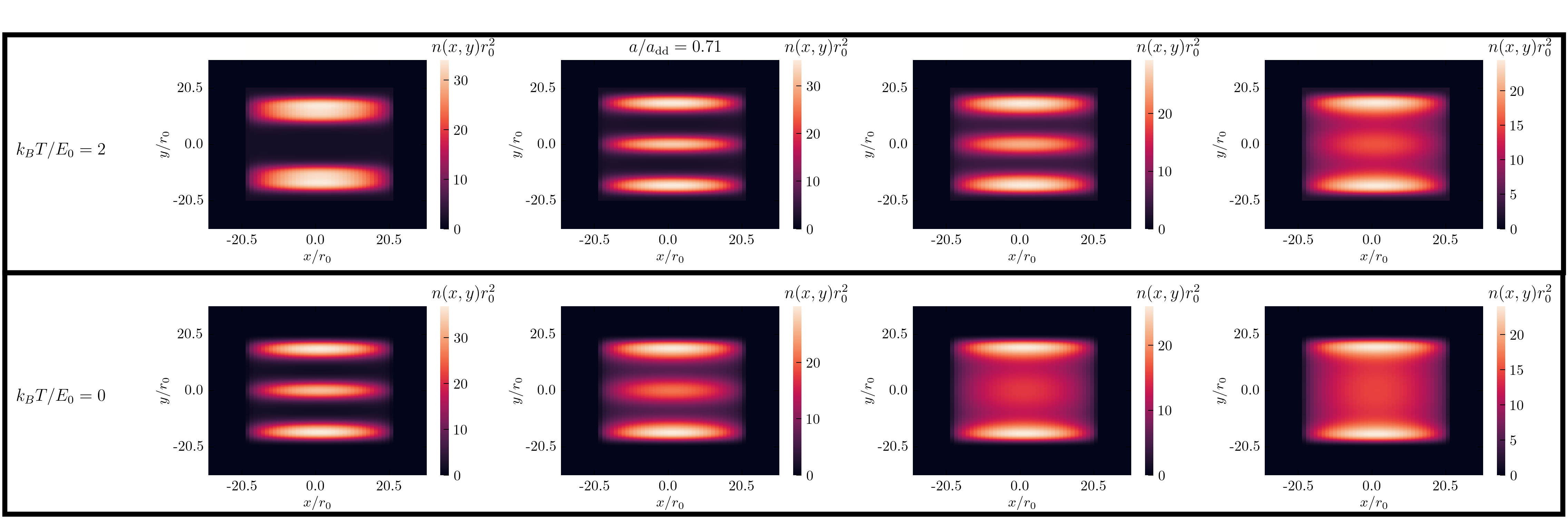}
\caption{Integrated total 2D density distribution ($n = \int dz \abs{\psi({\bf r})}^2 +  n_{\rm th}({\bf r}_{\perp})$) of the dipolar system at zero and finite temperature. In each row, scattering length $a$ is varied. The value of the parameters not indicated in the figure are the same as in the bottom row of Fig.~\ref{fig2}.}
\label{fig7}

\end{figure*}
\begin{figure*}[t]
\centering
\includegraphics[width=\linewidth]{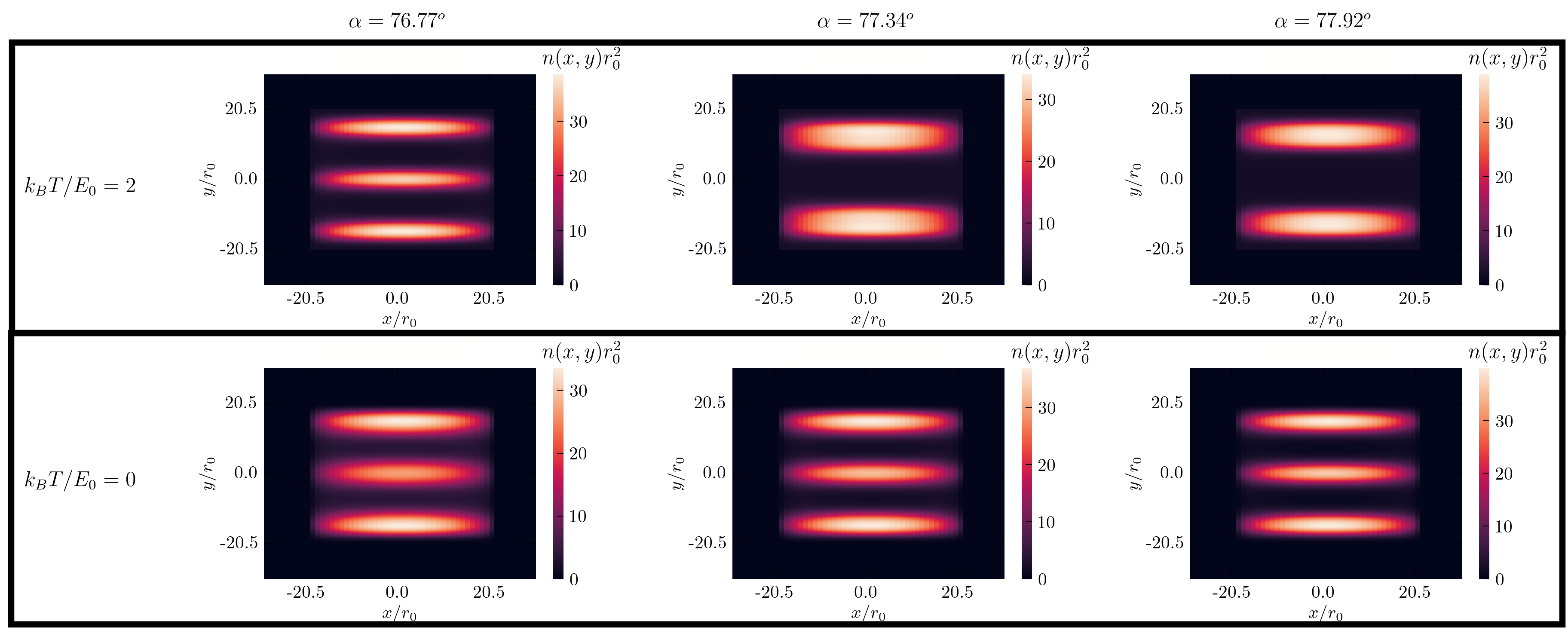}
\caption{Integrated total 2D density distribution ($n = \int dz \abs{\psi({\bf r})}^2 +  n_{\rm th}({\bf r}_{\perp})$) of the dipolar system at zero and finite temperature. In each row, the tilting angle $\alpha$ is varied. The value of the parameters not indicated in the figure are the same as in the bottom row of Fig.~\ref{fig2}.}
\label{fig8}
\end{figure*}

It must be remarked that, for all the temperatures considered in Fig.~\ref{fig5}, the condensate fraction remains above $f_c \geq 0.85$, meaning that, for the specific parameters considered, this striking change in the atomic density distribution happens despite the majority of atoms lying in the condensate. Compared to the system of Ref.~\cite{he2025:arxiv}, which features Erbium atoms and a BEC transition temperature of $T_c \sim 100$ nK, our temperature range in Fig.~\ref{fig5} for $^{166}$Er atoms surpasses significantly this threshold, while the condensate fraction remains small. This stems from the fact that we are employing a box trap instead of a harmonic trap. From a calculation of the BEC transition temperature for an ideal gas in 2D, it can be seen that a box trap is significantly more robust against thermal effects than a harmonic trap due to the different scaling of the energy levels with the quantum numbers $n_x$ and $n_y$: while the energy in a symmetric harmonic trap scales linearly with $n_x$ and $n_y$ ($E_{\rm h.o.} = \hbar \omega (n_x + n_y +1)$), the energy in the box trap scales quadratically ($E_{\rm box} = \frac{\hbar^2 \pi^2}{2 m L^2} \left( n_x^2 + n_y^2 \right)$). Thus, if $\hbar \omega$ is of the same order of magnitude as $\frac{\hbar^2 \pi^2}{2 m L^2}$, this implies that significantly less excited states are populated at the same temperature for the box trap. As an example, if one considers an ideal gas of $N=31000$ with the same atomic mass as $^{166}$Er and the harmonic traps of Ref.~\cite{he2025:arxiv} one obtains a BEC transition temperature of $T_{c,2D}^{\rm h.o.} = \left( \frac{6 N}{\pi^2} \right)^{1/2} \frac{\hbar \sqrt{\omega_x \omega_y}}{k_B} = 99.3$ nK, while a box trap with widths $L_x = L_y = 5.332$ $\mu$m (as in our case) yields $T_{c,2D}^{\rm box} = 2.7 \text{ } \mu$K~\cite{li2015:pra}.

The drastic change in the atomic distribution illustrated in Fig.~\ref{fig5} is amplified by the specific choice of parameters considered. This is because the impact of thermal effects is enhanced if the system is close to developing modulations at similar but smaller particle numbers at zero temperature (i.e. if the system is close to a structural transition). For instance, for the values of $a$, $\alpha$ and the box widths $L_x$ and $L_y$ of Fig.~\ref{fig5}, a zero temperature system develops a three-stripe configuration for $N \lesssim 27000$, which is smaller and close to the particle number of the figure, $N = 31000$. In contrast, we show in Fig.~\ref{fig6} the comparison between the total column density $n(y)$ at $k_B T / E_0 = 0$ and $k_B T / E_0 = 2.5$ for the same parameters as Fig.~\ref{fig5} except for the total atom number, which we set to $N=25000$ and $N=40000$. In both cases we can observe that finite temperature modifies the structure of the atomic cloud in a quantitative way only.

The results of Fig.~\ref{fig2} show that, at zero temperature, the 2D density in a given box trap changes upon varying the particle number, the tilting and the scattering length. For the same box trap as in Fig.~\ref{fig2}, which has widths $L_x/r_0 = L_y/r_0 = 41.02$, we study how the structural transition shifts if a finite temperature is considered. The results are shown in Figs.~\ref{fig7} and~\ref{fig8} where we compare the 2D density at two values of the temperature, $k_B T / E_0 = 0$ and $k_B T / E_0 = 2$, as we change the scattering length for a fixed tilting angle of $\alpha=77.35^o$, and the tilting angle for a fixed scattering length of $a/a_{\rm dd} = 0.7$, respectively. At $T = 0$, the particle number is fixed to $N=19935$, while at finite temperature it fluctuates between $N \in [19700,20000]$, which implies a neglectable fluctuation of $\Delta N/N \sim 1.1 \%$. The results show the same qualitative changes in the structure as the tilting and the scattering length are varied in each case at zero and finite temperature. However, the transition between the different structures is shifted by temperature to higher values of the scattering length and lower values of the tilting angle. This is because the thermal bunching effect previously discussed makes stripe configurations stable at a higher value of the scattering length. In regards to the tilting angle, this bunching effect stabilizes a lower number of higher density stripes at a lower tilting angle compared to the zero temperature case. Precisely, by comparing the results of two simulations with the same parameters except for temperature, we can see that given a modulated state, temperature can reduce the number of stripes that are formed. In regards to the aspect ratio of the box trap, we have checked that for a temperature of $k_B T / E_0 = 2$ the qualitative features shown in Fig.~\ref{fig3} at zero temperature remain at finite temperature.

The main conclusion to draw from the results in this Section is that, for the box traps considered, while there are specific parameter ranges where temperatures of the order $\sim 100 \text{ nK} \ll T_c$ can have a remarkable impact in the structure of the system by promoting modulations, this is not the case for the majority of the parameter space.

\section{\label{sec:conclusions}Conclusions}

To conclude, we have studied the equilibrium structure of a trapped, quasi two-dimensional system of tilted dipoles at zero and finite temperatures, restricting ourselves to the regime of large condensate fractions ($f_c \geq 0.8$). The quasi-2D nature of the system is due to a strong harmonic confinement along the $z$-axis. We consider a box trap confinement in the $x$-$y$ plane and dipoles polarized along a direction in the $x$-$z$ plane. The study of this system is motivated by the recent experimental observation of supersolid stripes in a similar configuration~\cite{he2025:arxiv}. We have adapted the finite temperature formalism followed in Refs.~\cite{oktel19,baena22} to our specific geometry. Unlike in these works, by explicitly considering the discrete excitations along the $z$-axis in the computation of the thermal fluctuations, we are able to obtain a convergent number of thermal atoms in our calculations and thus, estimate the condensate fraction in the dipole-dominated regime ($a_{\rm dd} > a$). At zero temperature, we have qualitatively characterized the behaviour of the trapped system when the different parameters (the particle number $N$, the scattering length $a$ and the tilting angle $\alpha$) are varied for a fixed box trap. We have also studied how changing the box trap aspect ratio can significantly change the behaviour of the system from liquid-like to gas-like as the width of the box along the $x$-axis, $L_x$, is decreased. At finite temperature, we have shown that an increase in temperature while the \textit{total} number of particles is kept fixed can promote modulations in the system, in qualitative agreement with the behaviour seen in recent Path Integral Monte Carlo simulations in a 3D tubular geometry~\cite{bombin25:prl}. We have also shown that, for the box traps considered, while temperatures of the order of $\sim 100$ nK can lead to drastic changes in the atomic density distribution for specific parameter ranges, this is not the case for the majority of the parameter space, where the atomic density is modified only quantitatively and not qualitatively by thermal effects.

Our results are helpful to understand and quantify the role played by finite temperature in typical experimental conditions in a field with recent emergent experimental activity~\cite{he2025:scienceadv,zhen2025:arxiv,he2025:arxiv,he2026:arxiv}. Specifically, the theory presented in this work can be used to obtain the thermal density of atoms in the supersolid regime and perform thermometry on a potential experiment in order to estimate the temperature or the atom number of the experimental supersolid samples. However, our results are restricted to temperatures where the condensate cloud remains majorly populated and allow us to study properties only at equilibrium. Future perspectives include the formulation of a real-time theory, where the real time extended Gross-Pitaevskii equation is solved simultaneously with the dynamics of the thermal excitations without resorting to the Popov approximation, which is inadequate in the dipole-dominated regime. This would make it possible to accurately simulate the parameter quenches performed in experiments. Moreover, the study of the system at equilibrium could be extended to higher temperatures by the application of c-field theories~\cite{Bisset2008:PRA,Bisset2009:PRA,pawlowski2013:PRA} to the dipolar system in the regime dominated by the dipole-dipole interaction, bypassing the strong limitation in the particle number present in exact finite temperature Monte Carlo methods.

\section{\label{sec:acknowledgements}Acknowledgements}

We thank Russell N. Bisset and Jordi Boronat for useful discussions. We acknowledge support by the Spanish Ministerio de Ciencia, Innovación y Universidades (grant PID2023-147469NB-C21, financed by MICIU/AEI/10.13039/501100011033 and FEDER-EU).

\appendix

\begin{figure}[b]
\centering
\includegraphics[width=\linewidth]{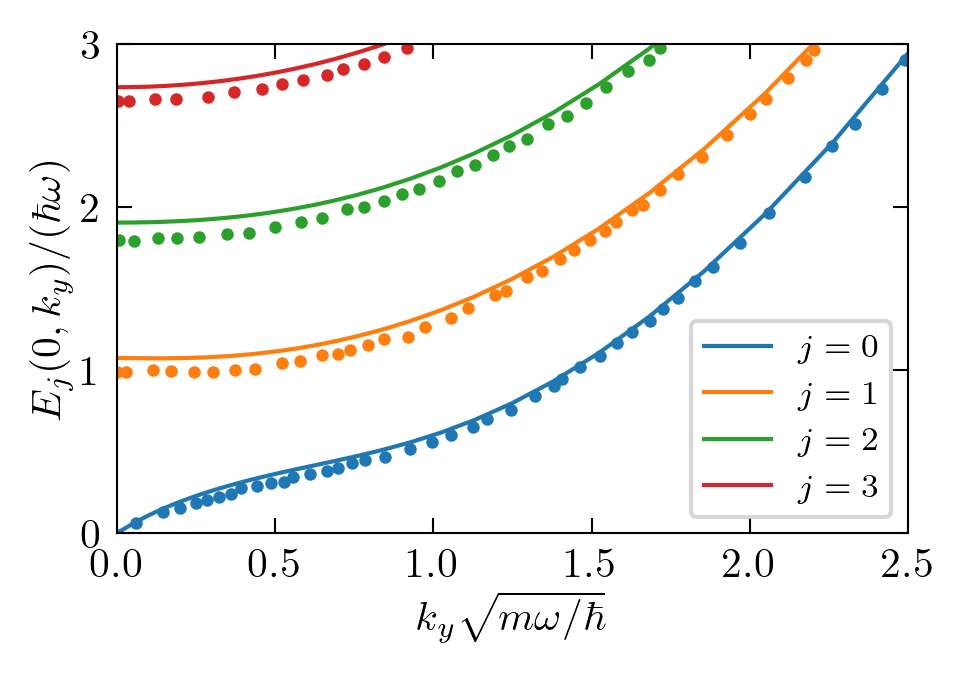}
\caption{Excitation spectrum of a quasi-2D, infinite dipolar system for $a/a_{\rm dd} = 0.875$, $n_0 r_0^2 = 24$, $\alpha = 60^o$ (same as Fig. 3a of Ref.~\cite{Baillie_2015}. The solid lines correspond to the solution of Eqs.~\ref{bdg1_approx}~\ref{bdg2_approx} while the dots correspond to the results of Ref.~\cite{Baillie_2015}. }
\label{figA1}
\end{figure}

\section{\label{sec:app_exc_spectrum}Benchmarking the approximation of Eqs.~\ref{bdg1_approx}~\ref{bdg2_approx} in the calculation of the excitation spectrum}

As mentioned in the main text, the thermal correction in Eq.~\ref{TeGPE} is obtained by solving Eqs.~\ref{bdg1_approx} and~\ref{bdg2_approx}, which correspond to the Bogoliubov-de Gennes equations after splitting the dependence of the Bogoliubov amplitudes as in Eqs.~\ref{ubg_approx} and~\ref{vbg_approx} and using the eigenstates of Eq.~\ref{GPE} as an ansatz to capture the dependence along $z$. In order to benchmark this approximation, we have compared the excitation spectrum obtained by solving Eqs.~\ref{bdg1_approx} and~\ref{bdg2_approx} with that of Fig. 3a of Ref.~\cite{Baillie_2015}, which is obtained by solving Eqs.~\ref{bdg1_full} and~\ref{bdg2_full}, i.e. without resorting into any splitting of the Bogoliubov amplitudes. The parameters of the calculation are $\alpha = 60^o$, $\hbar \omega/E_0 = 1$, $n_0 r_0^2 = 24$ and $a/a_{\rm dd} = 0.875$. We show the results in Fig.~\ref{figA1}. As we can see from the figure, the results for the excitation spectrum under the approximations considered in this work are in excellent agreement with the solution of the exact Eqs.~\ref{bdg1_full} and~\ref{bdg2_full}.

\begin{figure}[t]
\centering
\includegraphics[width=\linewidth]{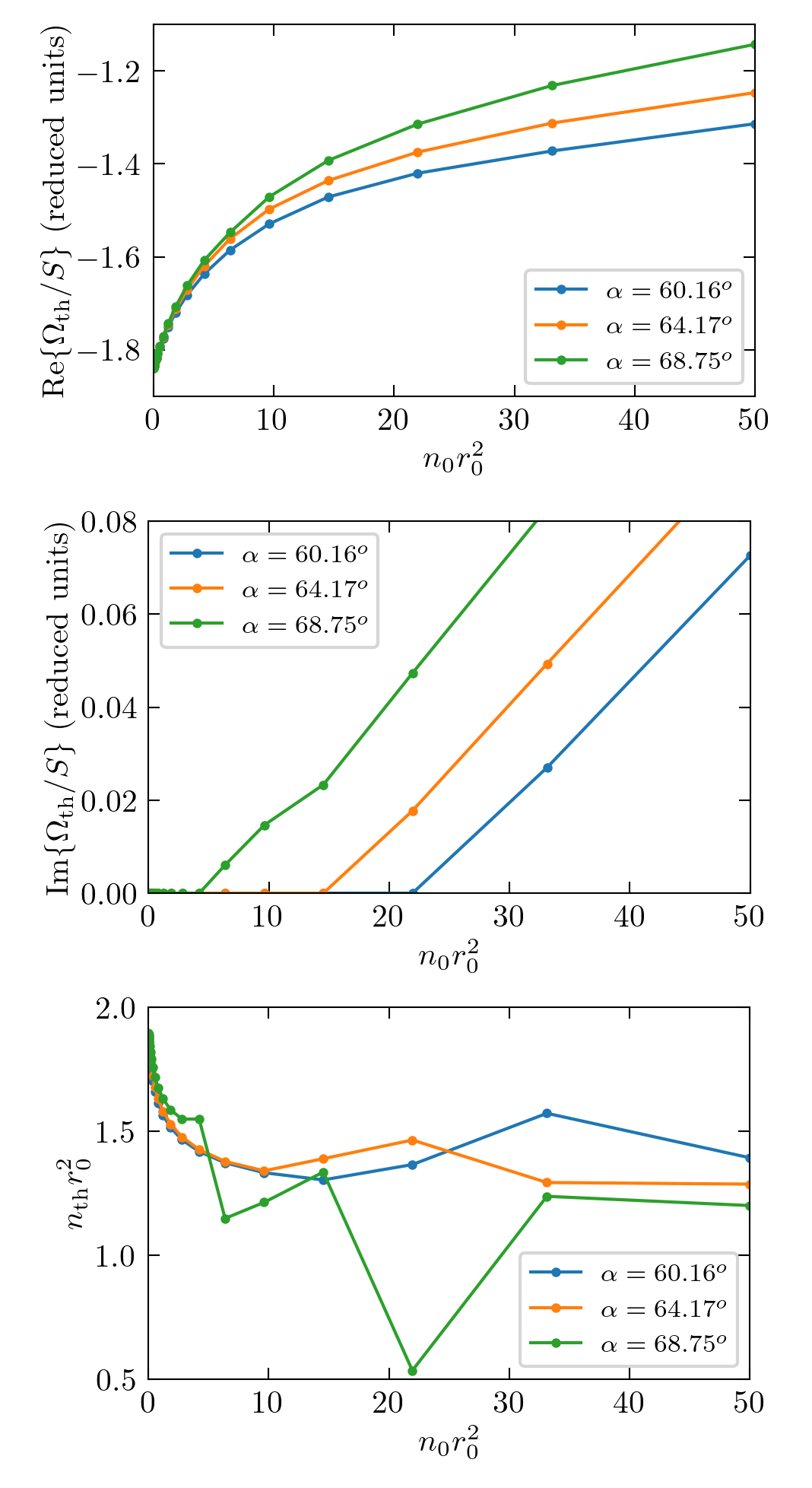}
\caption{Density functionals for the real and imaginary parts of $\Omega_{\rm th}/S (n_0)$ (top and center, respectively) and the real part of $n_{\rm th}(n_0)$ (bottom). The parameters for the calculations are $a/a_{\rm dd} = 0.7$, $\hbar \omega /E_0 = 1$, $L_x/r_0 = L_y/r_0 = 41.02$. }
\label{figB1}
\end{figure}

\section{\label{sec:app_f_t}Robustness of the results with respect to the thermal fluctuation parameters}

\begin{figure}[t]
\centering
\includegraphics[width=\linewidth]{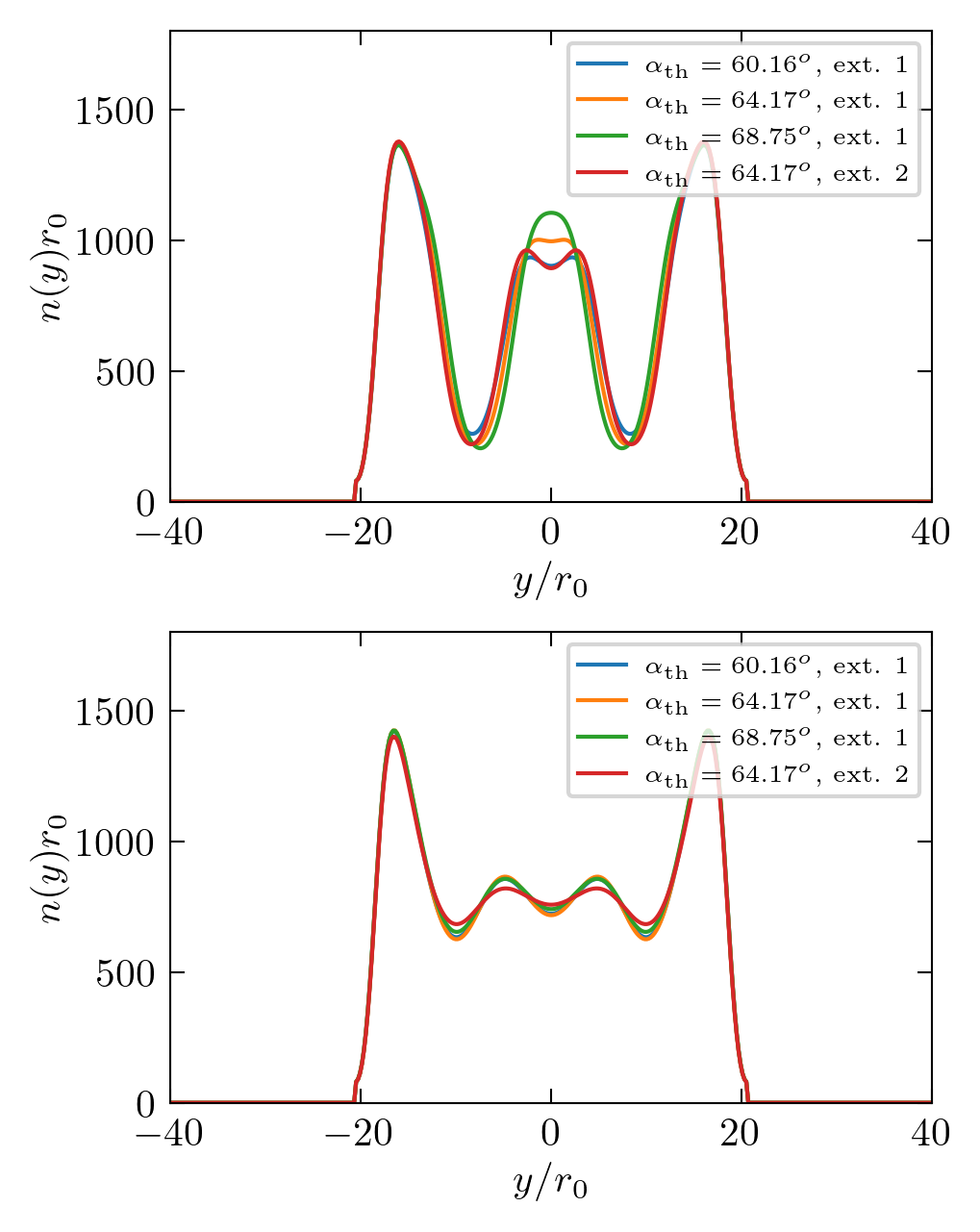}
\caption{ Total column density $n(y) = \int dx dz \text{ } \abs{\psi({\bf r})}^2 + \int dx \text{ } n_{\rm th}({\bf r}_{\perp})$ at $k_B T / E_0 = 2$ for $N=30317$ (top) and $N=34275$ (bottom). Different colors indicate a different value of the tilting $\alpha_{\rm th}$ employed in the calculation of the thermal fluctuations (see the text). ``ext. 1'' and ``ext. 2'' refer to two different density extrapolation prescriptions for the data of Eqs.~\ref{gc_th_co} and~\ref{n_th_co} (see the text). The rest of the parameters are the same as in Fig.~\ref{fig5}. }
\label{figB2}
\end{figure}

The excitation spectrum in Eq.~\ref{exc_spectrum} becomes imaginary at finite momenta for sufficiently large values of the tilting angle of the dipoles $\alpha$, and the 2D density of the condensate $n_0$. This is due to the well-known phenomenon of roton softening present in trapped dipolar systems~\cite{santos2003:prl}. Moreover, the excitation spectrum also develops a long wavelength instability if $V(0,0) < 0$, with $V({\bf k})$ the Fourier transform of the interaction (given in Eq.~\ref{V_FT}). These imaginary parts give rise to an unphysical complex contribution to the thermal part of the grand canonical potential as well as the density of thermal atoms, and even induce a numerical instability in the calculation of the latter. To illustrate this, we show in Fig.~\ref{figB1} the result of Eqs.~\ref{gc_th_co} and~\ref{n_th_co} as a function of $n_0$ for three distinct tilting angles: $\alpha = 60.16^o, 64.17^o$ and $68.75^o$. We employ a momentum cut-off of $k_x^{\rm co} r_0 = k_y^{\rm co} r_0 = 0.1326$, which corresponds to a box trap in the $x$-$y$ plane with $L_x/r_0 = L_y/r_0 = 41.02$. From the figure, we see that as the tilting angle increases, imaginary contributions arise in $\Omega_{\rm th}/S$ at a lower value of the 2D condensate density. We can also see that once $Im\{ \Omega_{\rm th}/S \} > 0$, the thermal density starts to drastically fluctuate, which is a consequence of a numerical instability.

In view of such results, it is preferable to avoid the imaginary contributions in the calculation of the thermal fluctuations, which imposes a limitation in the values of $\alpha$ and $n_0$ in the calculations. At the same time, we want to lay as close as possible to the relevant range of tilting angles, which corresponds to $\alpha > 68.75^o$. Thus, for the calculations presented in this work, we have chosen $\alpha = 64.17^o$ and a 2D condensate density range of $n_0 r_0^2 \in [0, 10]$. We denote by $n_{0,\text{max}}$ the upper bound of this range. This implies that Eqs.~\ref{gc_th_co} and~\ref{n_th_co} are computed for 2D condensate densities in the specified range and the subsequent data is afterwards fitted to the empirical functionals of Eqs.~\ref{fit_1} and~\ref{fit_2}, with the fitting parameters constituting the inputs to the TeGPE of Eq.~\ref{TeGPE}. It must be remarked that, as a consequence, the tilting angle employed in the calculation of the thermal fluctuations, which we now denote as $\alpha_{\rm th}$ and the one employed in the TeGPE, which we refer to as simply $\alpha$, are different. However, we must evaluate the effect of changing this setting in the solution of the TeGPE. Moreover, it is common that the local 2D condensate densities arising in the solution of the TeGPE lay above of the 2D density window where the data of Eqs.~\ref{gc_th_co} and~\ref{n_th_co} is computed and fitted, which constitutes an extrapolation approximation that must be also benchmarked. In order to do this, we have solved Eq.~\ref{TeGPE} computing the thermal fluctuations at different tilting angles and 2D condensate densities: ($\alpha_{\rm th. fluc} = 60.16^o$ ,$n_0 r_0^2 \in [0, 15]$), ($\alpha_{\rm th. fluc} = 64.17^o$ ,$n_0 r_0^2 \in [0, 10]$) and ($\alpha_{\rm th. fluc} = 68.75^o$ ,$n_0 r_0^2 \in [0, 4]$). We apply the extrapolation protocol mentioned above (which we refer to as Ext. 1), where the density functionals for the grand canonical thermal energy density and the density of thermal atoms that enter the TeGPE are simply given by Eqs.~\ref{fit_1} and~\ref{fit_2}. For the case with ($\alpha_{\rm th. fluc} = 64.17^o$ ,$n_0 r_0^2 \in [0, 10]$), we have also considered an alternative extrapolation prescription, referred to as Ext. 2, where we compute the density functionals as
\begin{align}
 \frac{\Omega_{\rm th}}{S}(n_0) =
 \begin{cases}
-a_1 -a_2 \exp \left(-a_3 n_0^{a_4}\right) & n_0 < n_{0,\text{max}} \\
  -a_1 -a_2 \exp \left(-a_3 n_{0,\text{max}}^{a_4}\right) & n_0 > n_{0,\text{max}}
 \end{cases}
 \\
 n_{\rm th}(n_0) =
 \begin{cases}
b_1 + b_2 \exp(-b_3 n_{0}^{b_4}) & n_0 < n_{0,\text{max}} \\
b_1 + b_2 \exp(-b_3 n_{0,\text{max}}^{b_4}) & n_0 > n_{0,\text{max}}
 \end{cases} \ .
\end{align}
We have considered a box trap of widths $L_x/r_0 = L_y/r_0 = 41.02$, a scattering length of $a/a_{\rm dd} = 0.7$, a tilting angle $\alpha = 77.35^o$ and particle numbers $N = 30317, 34275$. We show the results in Fig.~\ref{figB2}. From the figure we can see that the four prescriptions to compute the thermal fluctuations lead to very similar equilibrium column densities, thus indicating the robustness of our finite temperature calculations with respect to the specific choice of the tilting, the density range and the extrapolation to higher densities in the calculation of the thermal excitations.

\section{\label{sec:app_tech}Technical details of the numerical solution of the eGPE and the TeGPE}

Equations~\ref{eGPE} and~\ref{TeGPE} are numerically solved by imaginary time propagation. The number of points in each axis is $(N_x,N_y,N_z) = (75,400,40)$, while the imaginary time step in our units is of order $d\tau E_0/\hbar \sim 10^{-3}$. The number of iterations is of the order of $N_{\rm it} \sim 10^4$ or $N_{\rm it} \sim 10^5$, depending on the parameters. The covolution term with the DDI and the kinetic terms are computed through a Fast Fourier Transform (FFT) algorithm. For a given box trap of widths $(L_x, L_y)$, the size of the simulation box is chosen to be $(2L_x, 2L_y)$ while $L_z/r_0 = 20$, and we impose the condensate wave function $\psi({\bf r})$ and the thermal density $n_{\rm th}({\bf r})$ to be zero outside of the box trap. This guarantees that we are appropriately accounting for finite size effects, and improves on our previous implementation of the box trap in Ref.~\cite{baena2025:pra}, where the size of the box trap and the simulation box coincide and finite size effects were underestimated. Nevertheless, this underestimation does alter nor invalidate the conclusions obtained in that work.


\bibliography{paper_dipoles_tilt_finite_T}

\end{document}